\documentclass[preprintnumbers, floatfix, letterpaper, onecolumn,aps,prd,epsfig,nofootinbib,natbib,
longbibliography
]{revtex4-2}
\usepackage{bm,graphicx,dcolumn,epstopdf,epsf, latexsym,mathbbol, amssymb,amsmath,color,slashed, mathrsfs,mathcomp,simplewick}
\usepackage[center]{subfigure}
\usepackage{multirow}
\usepackage{graphicx}
\usepackage{makecell}
\usepackage{epstopdf}
\usepackage[table]{xcolor}
\usepackage[colorlinks,linkcolor=blue,citecolor=blue,urlcolor=blue]{hyperref}

\graphicspath{{figs/}} 

\begin{document}
\allowdisplaybreaks
 \newcommand{\bq}{\begin{equation}}
 \newcommand{\eq}{\end{equation}}
 \newcommand{\bqn}{\begin{eqnarray}}
 \newcommand{\eqn}{\end{eqnarray}}
 \newcommand{\nb}{\nonumber}
 \newcommand{\lb}{\label}
\newcommand{\f}{\frac}
\newcommand{\p}{\partial}
\newcommand{\PRL}{Phys. Rev. Lett.}
\newcommand{\PLB}{Phys. Lett. B}
\newcommand{\PRD}{Phys. Rev. D}
\newcommand{\CQG}{Class. Quantum Grav.}
\newcommand{\JCAP}{J. Cosmol. Astropart. Phys.}
\newcommand{\JHEP}{J. High. Energy. Phys.}
\newcommand{\orcid}[1]{\href{https://orcid.org/#1}{\includegraphics[width=10pt]{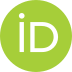}}}

\title{
  Non-Standard Thermal History and Formation of Primordial Black Holes in Einstein-Gauss-Bonnet Gravity}
\author{Yogesh\orcid{0000-0002-7638-3082}}
\email{yogeshjjmi@gmail.com}
\affiliation{Institute for Theoretical Physics and Cosmology,
Zhejiang University of Technology, Hangzhou, 310023, China}

\author{Abolhassan Mohmmadi\orcid{0000-0003-1228-9107}}
\email{abolhassanm@zjut.edu.cn; abolhassanm@gmail.com}
\affiliation{Institute for Theoretical Physics and Cosmology,
Zhejiang University of Technology, Hangzhou, 310023, China}


\begin{abstract}
Inflation provides a suitable environment for the formation of Primordial Black Holes(PBHs). In this article, we examine the formation of primordial black holes in the Mutated Hilltop inflation model coupled with the Einstein-Gauss-Bonnet term. A suitable choice of the coupling function with adjusted parameters can produce the USR regime during the inflationary phase, which lasts for some number of e-folds. The scalar field in this regime remains almost unchanged, and the first slow-roll parameter $\epsilon_1$ drops dramatically, leading to a significant enhancement to the curvature power spectrum for small scales so that it grows up to order of $\mathcal{O}(0.01)$; a crucial feature for producing PBH and secondary gravitational waves (GWs). We investigate the formation of PBHs for different sets of parameters. By considering the behavior of the scalar power spectrum, it is realized that the peak in the scalar power spectrum occurs in different scales. Then, the presented model is capable of predicting PBH formation in a wide range mass,  from $\mathcal{O}(10^{-14})M_{\odot}$ to $\mathcal{O}(10)M_{\odot}$, which are compatible with the LIGO-Virgo data. PBHs with mass $\mathcal{O}(10^{-5})M_{\odot}$ can account for the micro-lensing event in OGLE as well as the asteroid masses $\mathcal{O}(10^{-15})M_{\odot}-\mathcal{O}(10^{-12})M_{\odot}$ PBH can be attributed to $100\% $ dark matter present in the universe. The generated perturbations during inflation re-enter the horizons after the end of inflation, where the universe may be in a non-standard epoch rather than the radiation-dominant phase, with an equation of state $1/3< \omega \leq 1$. Due to this, the investigation is generalized to include the effect of non-standard thermal history on PBHs, and it is shown that this possibility can severely impact the formation and abundance of PBHs. The enhancement of the scalar power spectrum also leads to the scalar-induced GW, which is considered for different sets of parameters and the general value of the equation of state $\omega$. It is realized that by increasing $\omega$, there is a little enhancement of the induced GW; however, the peak of the induced GW is of the order of $10^{-8}$ for all cases. Moreover, produced secondary GWs are well within the detectable frequency ranges of current and future detectors such as NANOGrav, EPTA, PPTA, DECIGO, LISA, ET.

\end{abstract}

    \maketitle

\vspace{0.0001in}


\baselineskip=15.4pt

\section{Introduction\label{sec:intro}}
The inflationary paradigm is a remarkable and compelling paradigm that serves as one of the pillars of modern cosmology. Inflation is the name given to a brief period of accelerated expansion during the universe's first moments. Initially, inflation was employed to address the horizon and flatness problems, among other flaws in the Big Bang models \cite{Liddle:2000cg,1990eaun.book.....K,Guth:1980zm,Linde:1981mu,Mukhanov:1981xt,Sato:1981qmu,1996tyli.conf..771S,PhysRevLett.48.1220,Starobinsky:1982ee}. Later, it was realized that, inflation could also be responsible for large scale structure and formation of PBHs~\cite{PhysRevD.50.7173}. The idea of the PBHs was first discussed in 1967 by Zeldovich and Novikov in their seminal paper\cite{Zeldovich:1967lct}. Later, it was realized that PBHs with low mass could have been evaporated by the present epoch while heavier PBHs could be a candidate for the Dark Matter (DM) \cite{Hawking:1971ei,Carr:1974nx,Carr:1975qj}. The recent observations of the Black Hole Binary mergers from LIGO-Virgo surveys \cite{LIGOScientific:2016aoc,LIGOScientific:2016dsl,LIGOScientific:2016sjg,LIGOScientific:2016dsl,LIGOScientific:2016wyt,LIGOScientific:2017bnn,LIGOScientific:2017vox,LIGOScientific:2017ycc} hints towards the possibility of a primordial nature \cite{Fernandez:2019kyb} rather than the astrophysical origin. The mass of the PBHs can vary in a wide range from a few grams to a few solar masses; however, PBHs smaller than the $10^{15}~\rm gm$ could have been evaporated completely, whereas larger masses can account for  $100\% $ to few percent of total DM present in the universe \cite{Carr:2020gox}. Depending on the mass of PBHs, there are several observations that can impose constraints; small mass PBHs can be constraints through the galactic and extra-galactic $\gamma$-ray surveys, whereas lensing and the binary mergers impose constraints on heavy mass PBHs~\cite{Carr:2021bzv}. PBHs are formed in the early universe and a large number of mechanics can be accountable for their formation, such as the collapse of the density perturbations which are generated by a single/multi field inflation~\cite{PhysRevD.50.7173,Yokoyama:1998pt,Garcia-Bellido:2017mdw,Ballesteros:2017fsr,Hertzberg:2017dkh,Kinney:2005vj,Germani:2017bcs,Gangopadhyay:2021kmf,Pattison:2017mbe,Ezquiaga:2018gbw,Biagetti:2018pjj,Stewart:1997wg,Kohri:2007qn,Garcia-Bellido:1996mdl,Kawasaki:1997ju,Bhattacharya:2022fze,Lyth:2001nq,Kawasaki:2012wr,Kohri:2012yw,Yokoyama:1995ex,PhysRevLett.132.221003,Riotto:2023gpm,Riotto:2023hoz,Choudhury:2023vuj,Choudhury:2023jlt,Choudhury:2023rks,Bhattacharya:2023ysp,Sharma:2024whg,Gangopadhyay:2021kmf,Braglia:2020eai,Bhattacharya:2022fze}, 
from bubble collision \cite{Crawford:1982yz,Hawking:1982ga,La:1989st,PhysRevD.50.676,Sato:1980yn}, 
collapse of cosmic strings \cite{Planck:2013mgr,Blanco-Pillado:2017rnf,HAWKING1989237,PhysRevD.43.1106,HOGAN198487,Ghoshal:2023sfa,Datta:2024bqp,Borah:2023iqo} 
or domain walls \cite{Rubin:2001yw,PhysRevD.59.124014,Garriga:2015fdk,Deng:2016vzb,Liu:2019lul,Kopp:2010sh} etc. The relative abundance of PBHs as DM is determined by the mass spectrum of the PBH produced by each of these mechanisms and can be constrained by observations. In this work, we will focus on the PBHs formed through large perturbations in single-field inflationary models. PBH formed through the collapse of large overdensities is particularly intriguing since the primordial quantum fluctuations created during inflation can give rise to these overdensities \cite{Carr:1974nx,Carr:1975qj,Young:2019yug}. During inflation, scalar fluctuations are generated at all scales, and as the universe expands, the generated fluctuation exits the horizon. Once the inflation ends and the co-moving Hubble radius increases, these modes re-enter the horizon, become classical fluctuations and continue to grow. There are finite possibilities that they will gravitationally collapse and form PBHs if a significant amount of overdensities are present. The formation of PBH depends on the dominant component of the energy density at the time of the collapse. In the post-inflationary era, the mass and abundance of the PBHs depends on the inflation model and the dominant energy component of the universe at the time of collapse. In inflationary scenarios, when the small-scale scalar power spectrum ($\mathcal{P_{\zeta}}(k)$) is sufficiently large 
($\mathcal{P_{\zeta}}(k)_{PBH}/\mathcal{P_{\zeta}}(k)_{CMB}$) $\sim 10^{-7}$, it can lead to PBH formation after the end of inflation. When the PBHs are formed through the direct collapse of the overdensities resulting from the primordial inflationary fluctuations, we can employ a direct relation between the scale $1/k$, the scale at which the power spectrum $\mathcal{P}_\zeta(k)$ is sufficiently large, and the mass $(M_{PBH})$ of the produced PBHs. Further, the enhanced ($\mathcal{P_{\zeta}}(k)$) can be related to the mass function $\psi(M)$ with the abundance $f_{PBH}$ of the produced PBHs. However, when PBHs are formed in non-standard $1/3 < \omega \leq 1$ post-inflation epoch, $M_{PBH}$ and $f_{PBH}$ can significantly differ from the PBHs produced in the standard Radiation Dominated (RD) epoch \cite{Bhattacharya:2019bvk,Bhattacharya:2020lhc,Bhattacharya:2023ztw}. The required amplitude of $\mathcal{P_{\zeta}}(k)$ in non-radiation epochs requires the input of $\omega$ and radiation temperature $T_{RD}$. We present more details about this in the later sections.\\ 

Another intriguing avenue that goes hand in hand with PBHs is the Secondary Induced Gravitational Waves (SIGW). Both PBHs and SIGW require the enhancement in the primordial power spectrum ($\mathcal{P_{\zeta}}(k)$). Understanding the SIGW and PBHs can provide essential information about the universe at scales that other early universe observations, such as the Cosmic Microwave Background (CMB), cannot~\cite{Planck:2013mgr,Planck:2018jri}. In perturbation theory, the scalar and tensor perturbations are de-coupled at the first order, whereas in the second and higher orders, tensor and scalar modes are coupled. Then, an enhancement in $\mathcal{P_{\zeta}}(k)$ can source an enhancement in the tensor modes, which results in SIGWs~\cite{PhysRevD.75.123518,Baumann:2007zm,Kohri:2018awv,Espinosa:2018eve,Domenech:2021ztg}. Since SIGWs are sourced by $\mathcal{P_{\zeta}}(k)$, they have a primordial nature and appear to be stochastic. The growing interest in using gravitational waves (GWs) as a probe for the early universe holds promise for a complete understanding of primordial fluctuations, especially with the potential of ground- and space-based interferometric detectors and pulsar timing arrays. Therefore, PBH formation is closely linked to SIGWs, and both dynamics are critically dependent on the epoch of collapse and the era of SIGW sourcing from scalar modes, respectively. The frequency $f$ of the produced SIGW is related to the mode $k$ entering the horizon at the post-inflationary era when the GW is sourced.
 
In single field standard cold inflationary scenarios formation of PBHs and SIGW usually requires some exotic kind of features to be present in the inflation potential like a phase of ultra-slow roll~\cite{Garcia-Bellido:2017mdw,Ragavendra:2020sop,Ragavendra:2023ret}, a tiny bump or dip~\cite{Mishra:2019pzq}. These features are necessary to achieve sufficient enhancement to the primordial power spectrum required for the PBHs formation. This situation can be altered in the non-standard inflation pictures such as warm inflation~\cite{Arya:2019wck,Bastero-Gil:2021fac,Correa:2022ngq}, or modified gravity models of inflation, in which the enhancement in the power spectrum is achieved without exotic features being present in the inflationary potential. In this work we will explore the PBHs formation in Einstein-Gauss-Bonnet gravity where a simple inflationary potential combing with the suitable EGB coupling can result in the PBHs formation \cite{Kawai:2021edk,Ashrafzadeh:2024oll,Solbi:2024zhl,Ashrafzadeh:2023ndt,Zhang:2021rqs}. EGB gravity is a higher-dimensional theory that uses quadratic curvature adjustments to modify General Relativity. EGB gravity provides a rich environment to study non-standard inflationary dynamics when a coupling between a scalar field and the Gauss-Bonnet term is introduced~\cite{Kallosh:2013pby,1982PhLB..117..175S,Starobinsky:1983zz,Barvinsky:1994hx,Cervantes-Cota:1995ehs,Bezrukov:2007ep,Barvinsky:2008ia,DeSimone:2008ei,Gialamas:2020vto,Bezrukov:2008ej,Barvinsky:2009ii,Bezrukov:2010jz,Bezrukov:2013fka,Rubio:2018ogq,Koshelev:2020xby,Elizalde:2014xva,Gialamas:2023flv,Gialamas:2022xtt,Gialamas:2021enw,Gialamas:2020snr}. In EGB, adding the Gauss Bonnet term to the Hilbert-Einstein actions does not effect the total derivative of the equation of motions. The EGB coupling function $\xi(\phi)$ becomes dynamically essential when paired with a scalar field $\phi$, so that the Gauss-Bonnet term acts as a quantum correction to the Einstein-Hilbert action in string theory. In the past a large number of inflationary models have been explored in this modified gravity framework~\cite{vandeBruck:2015gjd,Guo:2009uk,Guo:2010jr,Koh:2016abf,Pozdeeva:2020shl,Satoh:2008ck,Jiang:2013gza,Koh:2014bka,Koh:2018qcy,Mathew:2016anx,Mathew:2016anx,Pozdeeva:2020apf,Pozdeeva:2016cja,Nozari:2017rta,Armaleo:2017lgr,Yi:2018gse,Yi:2018dhl,Odintsov:2018zhw,Fomin:2019yls,Fomin:2020hfh,Kleidis:2019ywv,Rashidi:2020wwg,Odintsov:2020sqy,Odintsov:2020zkl,Kawai:2021bye,Kawai:2017kqt,Oikonomou:2022xoq,Oikonomou:2022ksx,Cognola:2006sp,Odintsov:2020xji,Odintsov:2020mkz,Oikonomou:2020sij,Nojiri:2019dwl,Fomin:2019yls,Ashrafzadeh:2024oll,Solbi:2024zhl,Ashrafzadeh:2023ndt,Oikonomou:2024etl,Oikonomou:2024jqv,Odintsov:2023weg,Kawai:2023nqs,Kawai:1999pw,Kawai:1998ab,Nojiri:2024hau,Nojiri:2024zab,Elizalde:2023rds,Nojiri:2023mvi,Odintsov:2023aaw,Odintsov:2022rok,Odintsov:2022rok,Odintsov:2021urx,Pozdeeva:2020apf,Yogesh:2024mpa,Khan:2022odn,Pozdeeva:2024kzb}. The most widely use EGB coupling is the inverse function of the scalar field~\cite{Guo:2009uk,Guo:2010jr,Koh:2016abf,Pozdeeva:2020shl,Jiang:2013gza,Yi:2018gse,Odintsov:2018zhw,Kleidis:2019ywv,Rashidi:2020wwg}. In this paper, we will explore the formation of PBHs and SIGW within the framework of EGB, taking the Mutated Hilltop potential as our working model. \\

The rest of the paper is organized as follows: In section~\ref{EGB_model}, we will go over some
of the fundamental features in the EGB framework, and review the dynamics of the inflationary phase and resulting cosmological perturbations. In section~\ref{analysis}, we review the formation of PBHs in the general $\omega$ depending on the universe. In section~\ref{numerical_simulation}, we present the background analysis in the EGB domain using the Mutated Hilltop model. Section \ref{numerical_pbh_formation} shows the abundance of the PBHs in different epochs of the universe. In sections \ref{numerical_GW}, we analyze the SIGW for our model. Lastly, in section~\ref{conclusion}, we conclude with our conclusions and future work direction.

\section{Review of Einstein-Gauss Bonnet Gravity}
\label{EGB_model}

We investigate the EGB gravity theory with the following action \cite{Guo:2010jr,Jiang:2013gza,Koh:2014bka,Gao:2020cvb,Odintsov:2020sqy,Odintsov:2020xji,Rashidi:2020wwg,Azizi:2022yby,Khan:2022odn,Gangopadhyay:2022vgh,Nojiri:2023mvi,Odintsov:2023aaw,Nojiri:2023jtf}
\begin{equation}
    S = \int {\rm d}^4 x \sqrt{-g} \left[\frac{M_p^{2}}{2}R
	-\frac{1}{2}g^{\mu \nu}\nabla_{\mu}\phi
	\nabla_{\nu}\phi-V(\phi)-\frac{\xi(\phi)}{2} \mathcal{G}\right],
	\label{action1}
\end{equation}
so that $M_p$ is the reduced Planck mass, $g$ is the determinant of the metric $g_{\mu\nu}$, $R = g^{\mu\nu} R_{\mu\nu}$ is the Ricci scalar, $\phi$ is the scalar field with the potential $V(\phi)$, $\xi(\phi)$ is the coupling term which is a function of the scalar field, and $\mathcal{G}$ is the scalar invariant GB term defined as 
\begin{equation}
    \mathcal{G} = R_{\mu\nu\rho\sigma}R^{\mu\nu\rho\sigma}-4R_{\mu\nu}R^{\mu\nu}+R^2.
\end{equation}
The universe is assumed to be described by a spatially flat FLRW metric given by
\begin{equation}
    {\rm d}s^2=-{\rm d}t^2+a^2(t) \delta_{ij}{\rm d}x^i {\rm d}x^j \,
\end{equation}
where $a(t)$ is the scale factor. Taking variation of the action with respect to the metric and imposing the metric leads to the Friedmann equations as \cite{Guo:2010jr,Jiang:2013gza,Koh:2014bka,Gao:2020cvb,Odintsov:2020sqy,Odintsov:2020xji,Rashidi:2020wwg,Azizi:2022yby,Khan:2022odn,Gangopadhyay:2022vgh,Nojiri:2023mvi,Odintsov:2023aaw,Nojiri:2023jtf}
\begin{eqnarray}
    3 M_p^{2}H^2 &=& \frac12\dot{\phi}^2+V(\phi)+12\dot{\xi}H^3, \label{Friedmann} \\ 
    -2 M_p^{2} \dot{H} &=& \dot{\phi}^2 - 4\ddot{\xi}H^2 - 4\dot{\xi}H\left(2\dot{H} - H^2\right)\, \label{Friedmann2}
\end{eqnarray}
and by taking the variation of the action with respect to $\phi$, the equation of motion of the scalar field is achieved as
\begin{eqnarray}\label{eom}
    \ddot{\phi}+3H\dot{\phi} = - V_{,\phi} - 12\xi_{,\phi}H^2\left(\dot{H}+H^2\right),
\end{eqnarray}
here, the dot stands for the time derivative. The slow-roll approximations are exhibited by the usual slow-roll parameters that we have in standard cosmology as
\begin{equation}
    \epsilon_1 \equiv \frac{-\dot{H}}{H^2}, \quad 
    \epsilon_2 \equiv \frac{\ddot{\phi}}{H \dot{\phi}}, \quad
\end{equation}
so that the smallness of these parameters guarantees a quasi-de Sitter expansion phase that lasts long enough. In addition, there are two other slow-roll parameters in EGB gravity theory, given by
\begin{equation}
    \delta_1 \equiv 4 \dot{\xi} H, \quad 
    \delta_2 \equiv \frac{\dot{\delta}_1}{H \delta_1}.
\end{equation}
The dynamical equations of the model are simplified by applying the slow-roll approximations, $\epsilon_i \ll 1$ and $\delta_i \ll 1$, that can be rewritten as
\begin{eqnarray}
    && H^2 \approx \frac{V(\phi)}{3 M_p^2}, \label{Friedmann1_sr} \\
    && -2 M_p^2 \dot{H} \approx \dot{\phi}^2 + 4 \dot{\xi} H^3, \label{Friedmann2_sr} \\
    && 3 H \dot{\phi} + V_{,\phi}(\phi) \approx - 12 \xi_{,\phi}(\phi) H^4 \label{eom_sr}
\end{eqnarray}
The cosmological perturbations generated during the inflationary phase are stretched outside the horizon, and then they re-enter the horizon after inflation. Under the slow-roll conditions, the power spectrum related to the scalar and tensor perturbations at the sound horizon crossing, $c_{s,t} k=a H$, are given respectively by \cite{Horndeski:1974wa,DeFelice:2011zh,Kawaguchi:2022nku,Jiang:2013gza,Koh:2014bka,Odintsov:2020zkl} 
\begin{equation}\label{Ps}
    \mathcal{P}_{\zeta}(k) = \frac{H^2}{8\pi^2 Q_s c_s^3}\biggr|_{c_s k=a H}\,, \qquad 
    \mathcal{P}_t=\frac{H^2}{2\pi^2 Q_t c_t^3}\biggr|_{c_t k=a H}
\end{equation}
in which the introduced parameters are defined as 
\begin{eqnarray}
    Q_s & = & 16\frac{\Xi}{\Delta^2}Q_t^2+12Q_t\, \label{Qs} \\
    c_s^2 & = & \frac{1}{Q_s}\left[ \frac{16}{a}\frac{d}{d t}
	\left(\frac{a}{\Delta}Q_t^2\right)-4c_t^2 Q_t\right]\, \label{cs2} \\
    Q_t & = & \frac{1}{4}\left(-4H\xi_{,\phi}\dot{\phi}+M_p^2\right)\, \label{Qt} \\ 
    c_t^2 & = & \frac{1}{4Q_t}\left(M_p^2-4\xi_{,\phi\phi}\dot{\phi}^2-4\xi_{,\phi}\ddot{\phi}\right), \\
    \Xi & = & \frac{1}{2}\dot{\phi}^2-3 M_p^2 H^2+24 H^3 \xi_{,\phi}\dot{\phi}\, \\
    \Delta & = & M_p^2 H-6 H^2 \xi_\phi\dot{\phi}, 
\end{eqnarray}
Note that by taking $\xi = 0$, the background dynamical equations and the power spectra return to the corresponding ones in the standard cosmology. The possibility of the ghost and Laplacian instabilities occurrence can be avoided if the following conditions are satisfied
\begin{equation}\label{ghost_conditions}
    Q_s>0, \qquad c_s^2>0, \qquad Q_t>0,\qquad c_t^2>0 .
\end{equation}
The scalar spectral index is defined through the scalar power spectrum, which, by applying the slow-roll conditions, is obtained as \cite{DeFelice:2011zh,Kawaguchi:2022nku,Jiang:2013gza} 
\begin{equation}\label{ns}
    n_s-1 \equiv \frac{d \ln {\cal P}_{\zeta}}{d \ln k}\biggr|_{c_s k=aH} \simeq -2\epsilon_1-\frac{2\epsilon_1 \epsilon_2 - \delta_1 \delta_2}{2\epsilon_1-\delta_1}\,.
\end{equation}
Also, the tensor-to-scalar ratio is obtained as
\begin{eqnarray}\label{r}
    r \equiv \frac{{\cal P}_t}{{\cal P}_{\zeta}}\simeq 8\left(2\epsilon_1-\delta_1\right).
\end{eqnarray}
According to the latest data release of Planck-2018 TT, TE, EE + lowE + lensing + BK15 + BAO, the amplitude of the scalar power spectrum at the pivot scale $k_\star = 0.05 {\rm Mpc^-1}$ is $\mathcal{P}_{\zeta}(k_\star) = 2.098 \times 10^{-9}$ ~\cite{Planck:2018jri}. The scalar spectral index is reported as $n_s=0.9649\pm 0.0042 \; (68\,\%\,{\rm CL})$ and tensor-to-scalar ratio is bounded as $r<0.058 \; (95\%~ {\rm CL})$ ~\cite{Planck:2018jri}, and the bound gets tighter as $r<0.038 \; (95\%~ {\rm CL})$, according to Planck-2018 TT, TE, EE + lowE + lensing + BK18 + BAO ~\cite{Campeti:2022vom}. \\

\subsection{Inflationary phase and power spectrum enhancement}
The potential of the scalar field is considered to be Mutated Hilltop potential, given by ~\cite{Pal:2009sd,Pinhero:2019sbg,Yogesh:2024zwi,Gangopadhyay:2022vgh}
\begin{equation}
    V(\phi) = V_0 \; \big( 1 - \rm sech(\alpha \; \phi) \big),
\end{equation}
where $\alpha$ is a free constant parameter. The coupling function $\xi(\phi)$ is considered as~\cite{Kawai:2021edk,Ashrafzadeh:2023ndt,Zhang:2021rqs}
\begin{equation}\label{coupling}
    \xi(\phi) = \xi_0 \; \tanh\big[\xi_1 (\phi - \phi_c) \big],
\end{equation}
with three free constant parameters $\xi_0$, $\xi_1$, and $\phi_c$. \\ 
The formation of PBH and the scalar-induced GWs necessitate a high enhancement in the magnitude of the scalar power spectrum on small scales. Such an enhancement can be acquired by having a short USR phase within the inflationary phase. Although in Einstein's gravity, USR is achieved by adding an extra term to the potential, one can produce the same phase by choosing a proper coupling function with adjusted parameters in EGB gravity theory. There are two criteria required to be satisfied by the model. First, The model must be consistent with the Planck data at the pivot scale. Second, an enhancement of the power spectrum by about $10^7$ order of magnitude at small scales. To achieve these aims, the free parameters of the model must be well-adjusted. For the above-chosen coupling function, the parameter $\phi_c$ determines the position of the produced peak in the scalar power spectrum, while $\xi_0$ and $\xi_1$ specify the height and width of the peak. Moreover, utilizing the observational value for the scalar power spectrum at the pivot scale, one can constrain the value of the free parameter $V_0$. \\
The USR phase is achieved as the scalar field crosses near a fixed point, where the parameters $\dot{H}$, $\dot{\phi}$, and $\ddot{\phi}$ vanish near this fixed point. Applying these conditions on Eqs.\eqref{eom} results in 
\begin{equation}\label{x0_constraint}
    \left( V_{,\phi}+\frac{4 \xi_{,\phi}{V(\phi)}^2}{3 M_p^4} \right)
	\Biggr|_{\phi=\phi_c} = 0\,.
\end{equation}
This is the main relation to determine the $\xi_0$ parameter of the model. There is a USR regime in the neighborhood of the fixed point, where the slow-roll approximations are no longer valid. Therefore, Eq.\eqref{Ps} can not be used to measure the power spectrum of the scalar perturbations. \\
To calculate the power spectrum, the Mukhanov-Sasaki should be solved numerically. In this order, the background equations are first required to be solved. Eqs.\eqref{Friedmann}-\eqref{eom} are rewritten in terms of the number of e-fold $N$, using $dN = H dt$. The inflationary phase is assumed to last for about $N = 60$ e-fold of expansion, where the horizon crossing occurs at $N = 0$. The initial conditions are determined through the dynamical equations of the model in the slow-roll regime. After solving the background equations, the solution is fed to the Mukhanov-Sasaki equation, given by \label{Kawaguchi:2022nku}
\begin{equation}\label{MS}
    \nu_k''+\left( k^2-\frac{z_s''}{z_s}\right)\nu_k=0\,,
\end{equation}
where $\nu_k=z_s \zeta_k$ and $z_s=a \sqrt{2Q_s c_s}\,$, and prime indicates derivative with respect to the conformal time $a d\tau_s = c_s dt$. The initial conditions for the perturbations are set at the subhorizon scales, i.e., $c_s k \ll a H$, where the modes are well inside the horizon. Then, the Bunch-Davies vacuum state could be a good choice for the initial condition as
\begin{equation}\label{nu_ic}
    \nu_k=\frac{1}{\sqrt{2 k}}e^{-ik\tau_s}\,.
\end{equation}
The evolution of the curvature fluctuations $\zeta_k$ is determined by the Mukhanov-Sasaki equations. They evolve from the subhorizon scales, cross the horizon, and reach the superhorizon scales, where they freeze. Using the solution of the Mukhanov-Sasaki equation, the scalar power spectrum is precisely calculated as a function of the mode $k$ by
\begin{equation}
    \mathcal{P}_{\zeta}(k)\equiv \frac{k^3}{2\pi^2} \left| 
       \zeta_k (\tau_{s},{ k})\right|^2\,.
\end{equation}
Calculating the power spectrum will be the first step for considering the formation of PBH and generation of the scalar-induced GWs, which will be investigated in the subsequent sections. In the following, we consider two different values of the parameter $\alpha$, and for each case, we present four sets of the parameters.

\section{Non-Standard Thermal History and PBH Formation}
\label{analysis}
In standard cold-inflationary scenarios, after the end of inflation, we require a reheating phase to re-populate the universe and enter the radiation era. However, it is not possible to constrain the reheating temperature precisely as it depends heavily on the inflation model. The reheating temperature can vary in vast order from the end of inflation ($ \mathcal{O} 10^{16} \rm Gev$) to the onset of Big Bang Nucleosynthesis (BBN) ($ \sim 5 \rm Mev$). The observation from the abundance of light elements requires that the radiation epoch should last up to the onset of BBN.  So, there is a finite possibility for the universe to transit into the phases where  $\omega \neq 1/3$ between the end of inflation and the start of BBN. Since the physics of PBH formation crucially depends on the epoch/time when the over-densities collapse to form a PBH, it is important to consider the effect of non-standard post-inflationary epochs on the formation of PBHs. It is certainly possible that the post-reheating era is dominated by the particle species $\Phi$ with $\rho_{\Phi} \propto a^{-3 (1+ \omega)}$. Such as an early matter-dominated (EMD) era $\omega=0$ moduli fields in string theory models can give rise to EMD~\cite{Bhattacharya:2021wnk,Allahverdi:2020bys,Vilenkin:1982wt,Coughlan:1983ci,Starobinsky:1994bd} or kinetic energy dominated epoch $\omega \approx 1$, which generally happens in quintessential inflation models~\cite{Peebles:1998qn,Ahmad:2019jbm}, or an era where $\omega < 1/3$ for a small duration. There could be other scenarios where $\omega$ can take more general values. However, in this paper, we are interested in exploring the effect of stiff EoS $1/3< \omega \leq 1$~\cite{DiMarco:2018bnw} on the production of PBH and SIGW. 
Because the overdensity, which is defined as $\delta \equiv \frac{\rho - \bar{\rho}}{\rho}$ where $\rho$ indicates the local densities and $\bar{\rho}$ stands for the mean densities, are significantly large in the extreme tails of the probability distribution, it would be a rare event to generate PBHs from large-scale density fluctuations. When the overdensities $\delta$ are larger than a specific threshold $\delta_c$, i.e., $\delta > \delta _c$, fluctuations can collapse and form a PBH. $\delta _c$ depends on the background equation of state (EOS) $\omega$, which can impact the mass and abundance of the produced PBHs. Formation of PBHs in the general $\omega$ dependent epoch was first discussed in the \cite{Bhattacharya:2019bvk}; however, in this paper, we closely follow the prescription provided in the recent review~\cite{Bhattacharya:2023ztw}. 

The scale $k$ is related to the mass $M$ of the produced PBHs. In the $\omega$ dependent epoch, the Hubble parameter is proportional to $H\propto a^{-3(1+w)/2}$. As the scale $k$ re-enter the horizon, $k =a H$, one has
\begin{equation}
k\propto H^{\frac{1+3w}{3(1+w)}}.
\label{kH1}
\end{equation}
A horizon of size $H^{-1}$ can have the total mass $M_{H}=\frac{4\pi H^{-3}\rho}{3}$, and usually a small amount of this mass collapsed to form PBHs\footnote{Typically assumed to be $\gamma =0.33$, although, it can depend on the epoch of formation~\cite{PhysRevLett.74.5170}.}: $M=\gamma M_{H}$. From the Friedman equation $H^2 = \frac{\rho}{3M_P^2}$ one can write the Mass and Hubble parameter, as
\begin{equation}
M=\frac{4\pi \gamma M_P^2}{H}.
\label{MassHub}
\end{equation}
Thus, the relation between $M$ and $k$ can be written as:
\begin{equation}
M\propto \bigg(\frac{k}{4\pi \gamma M_P^2}\bigg)^{\frac{1+3w}{3(1+w)}}.
\label{Mkapprox}
\end{equation}
Accepting the matching relation, the exact dependence of $k$ and $M(k)$ through the temperature $T$ can be expressed as:
\begin{equation}
H(T)=\frac{H(T)}{H(T_{\rm RD})}H(T_{\rm RD})=\bigg(\frac{a(T)}{a(T_{\rm RD})}\bigg)^{-\frac{3(1+w)}{2}}\bigg(\frac{\pi ^2 g_*(T_{\rm RD})}{45M_P^2}\bigg)^{1/2}T_{\rm RD}^2,
\end{equation}
in which, we take $\rho (T_{\rm RD})=\rho _{R}(T_{\rm RD})+\rho _w{T_{\rm RD}}=2\rho (T_{\rm RD})=2\frac{\pi ^2}{30}g_*(T_{\rm RD})T_{\rm RD}^4$. Also, $g_*(T)$ and $g_s(T)$ denote the energy and entropy degrees of freedom, respectively. Admitting the entropy conservation, we can write $g_s(T)a(T)^3T^3$ at every epoch, and from $k=a(T)H(T)$ we obtain:
\begin{equation}
k=\bigg(\frac{\pi ^2 g_*(T_{\rm RD})}{45M_P^2}\bigg)^{1/2} a_{\rm eq}T_{\rm eq} \bigg(\frac{g_s(T)}{g_s(T_{\rm RD})}\bigg)^{\frac{1+w}{2}}\bigg(\frac{g_s(T_{\rm eq})}{g_s(T)}\bigg)^{\frac{1}{3}}T^{\frac{1+3w}{2}}T_{\rm RD}^{\frac{1-3w}{2}},
\label{relkT}
\end{equation}
here the subscript `eq' signifies the time of matter radiation equality. This leads to the following expression for $M(k)$:
\begin{equation}
M(k)=4\pi \gamma M_P^2 \bigg(\frac{\pi ^2 g_*(T_{\rm RD})}{45M_P^2}\bigg)^{\frac{1}{1+3w}} \bigg(\frac{g_s(T_{\rm eq})}{g_s(T_{\rm RD})}\bigg)^{\frac{1+w}{1+3w}}(a_{\rm eq}T_{\rm eq})^{\frac{3(1+w)}{1+3w}} T_{\rm RD}^{\frac{1-3w}{1+3w}}k^{-\frac{3(1+w)}{1+3w}}.
\label{Mkexact}
\end{equation}

The probability of the gravitational collapse of an overdensity $\delta$ correspondence to a PBH is $P(\delta)$. The probability that PBH of mass $M$ has formed is encoded in the mass fraction $\beta (M)$. For an $\omega$ dominated epoch, $\beta (M)$  crucially depends on $\omega$ via the critical overdensity $\delta _c(\omega)$. In the Press--Schechter formalism, we write,
\begin{equation}\label{eq:beta}
	\beta(M) = \int_{\delta_c}^{\infty} d\delta ~P(\delta).
\end{equation}
Assuming the Gaussian profile of the density fluctuations,
\begin{equation}
P(\delta)= \frac{2}{\sqrt{2\pi}\sigma(M)} \exp\left(-\frac{\delta^2}{\sigma(M)^2}\right),
\label{probdel}
\end{equation}
here, $\sigma (M)$ denotes the variance of the density fluctuation for a scale relating to PBH mass $M$ that can be related to the primordial curvature power spectrum $P_{\zeta}(k)$ as: 
\begin{equation}
	\sigma^2(k)=\frac{4(1+w)^2}{(5+3w)^2}\int \frac{dq}{q} W^2 \left(\frac{q}{k} \right) \left( \frac{q}{k} \right)^4 T^2 \left(\frac{q}{k}  \right) P_{\zeta}(q),\label{sigmaM1}
\end{equation}
where $T(q/k)= 3 \left (\rm Sin (\it l.\sqrt{\omega})-\it l.\sqrt{\omega} ~\rm Cos(\it l. \sqrt{\omega}) \right)/(\it l.\sqrt{\omega})^3$  is the $\omega$ depended transfer function with $l=q/k$  and  $W(q/k) = \exp (-q^2/k^2/2)$ denotes the window function that smoothens the perturbations. 

The fraction energy density that goes into the collapse of the PBH is  $\frac{\rho _{\rm PBH}}{\rho}\vert_{i}=\gamma \beta (M)$, where the subscript $i$ denotes the formation time of the PBH with mass $M$, and $\rho$ is the total energy density present in the universe at the time of formation. Now, PBH abundance (fraction of DM in the form of PBH), is defined through the PBH mass function $\psi (M)$ as 
\begin{equation}\label{eq:psi}
	\psi (M)=\frac{1}{M}\frac{\Omega_{\rm PBH}(M)}{\Omega _{\rm DM}}\bigg \vert _0\,.
\end{equation}
$\psi (M)$ carries the information of the fractional energy that goes into the PBH at formation. The present value of the mass function $\psi (M)$ can be determined by the energy evolution in PBH after formation until the epoch of matter-radiation equality (denoted with suffix `eq'). Since after formation, PBHs behave as standard matter, and they evolve similarly to the background energy density in the matter-dominated epoch we write,
\begin{eqnarray}
\psi (M)&=&\frac{1}{M}\frac{\Omega_{\rm PBH}(M)}{\Omega_c}=\frac{1}{M}\frac{\rho_{\rm PBH}(M)}{\rho_c}\bigg\vert_{\rm eq}\nonumber \\
&=&\frac{1}{M}\frac{\rho_{\rm PBH}(M)}{\rho_{\rm rad}}\bigg\vert_{\rm eq}\bigg(\frac{\Omega_m h^2}{\Omega_c h^2}\bigg)\nonumber \\
&=&\frac{1}{M}\frac{\rho_{\rm PBH}(M)}{\rho_{w}}\bigg\vert_{T_{\rm RD}}\bigg(\frac{a(T_{\rm eq})}{a(T_{\rm RD})}\bigg)\bigg(\frac{\Omega_m h^2}{\Omega_c h^2}\bigg)\nonumber \\
&=&\frac{1}{M}\frac{\rho_{\rm PBH}(M)}{\rho_{w}}\bigg\vert_{T}\bigg(\frac{a(T_{\rm RD})}{a(T)}\bigg)^{3w}\bigg(\frac{a(T_{\rm eq})}{a(T_{\rm RD})}\bigg)\bigg(\frac{\Omega_m h^2}{\Omega_c h^2}\bigg)\nonumber \\
&=& \frac{\gamma \beta(M)}{M}\bigg(\frac{g_s(T_{\rm RD})}{g_s(T)}\bigg)^{-w}\bigg(\frac{g_s(T_{\rm eq})}{g_s(T_{\rm RD})}\bigg)^{-1/3}\bigg(\frac{T}{T_{\rm RD}}\bigg)^{3w}\bigg(\frac{T_{\rm RD}}{T_{\rm eq}}\bigg)\bigg(\frac{\Omega_m h^2}{\Omega_c h^2}\bigg).
\label{psiM0}
\end{eqnarray}
Here, in the second line, we admit that the matter and radiation energy density are equal at $T_{\rm eq}$. Similarly, we have applied the equality of the radiation energy density and the species' energy density with EoS $w$ at $T_{\rm RD}$ in the third line. Lastly, using the entropy conservation, Eq.~\eqref{relkT} and Eq.~\eqref{Mkexact}, the mass function can be formulated in terms of the PBH mass $M$ as: \footnote{In our analysis we take  $g_*(T_{\rm RD})=g_s(T_{\rm RD})=106.75$ when $T_{RD}= 100 ~\rm GeV$, $g_s(T_{\rm eq})=3.38$, $\Omega_m h^2 = 0.14237$ and $\Omega_c h^2 = 0.1200$ }
\begin{equation}
\psi (M)=\frac{\gamma}{T_{\rm eq}}(4\pi \gamma M_P^2)^{\frac{2w}{1+w}}\bigg(\frac{g_s(T_{\rm RD})}{g_s(T_{\rm eq})}\bigg)^{1/3}\bigg(\frac{\pi ^2 g_*(T_{\rm RD})}{45M_P^2}\bigg)^{-\frac{w}{1+w}}\bigg(\frac{\Omega_m h^2}{\Omega_c h^2}\bigg)T_{\rm RD}^{\frac{1-3w}{1+w}}\beta (M)M^{-\frac{1+3w}{1+w}}.
\label{MpsiMw}
\end{equation}
Now we can evaluate the total contribution of PBH to the DM as:
\begin{equation}
f_{\rm PBH}=\int dM \psi (M).
\label{ftot}
\end{equation}
This prescription is valid for a general $\omega$ dominated epoch. However, in this article, we restrict ourselves to $1/3< \omega \leq 1$. It is evident from Eq.~\ref{MpsiMw} that $\psi (M)$ is a function of $\omega$ and $T_{RD}$. Then, different values of $\omega$ and $T_{RD}$ will lead to drastically different values of $\psi (M)$. This dependency will also be reflected in the total DM fraction ($f_{\rm PBH}$). We will present the detailed analysis in the later sections.

\section{Numerical solutions}\label{numerical_simulation}
In this section, the numerical solutions of the model are discussed for two different values of $\alpha$. Following Sec.II, Eqs.\eqref{Friedmann}-\eqref{eom} and Mukhanov-Sasaki equations are solved numerically for different sets of $(\phi_c, \xi_0, \xi_1)$ parameters. In the subsequent subsections, we study the background and perturbations solution in detail.

\subsection{$\alpha = 1$}
\begin{figure}[t]
    \centering
    \subfigure[\label{a1_bgn_phi}]{\includegraphics[width=7cm]{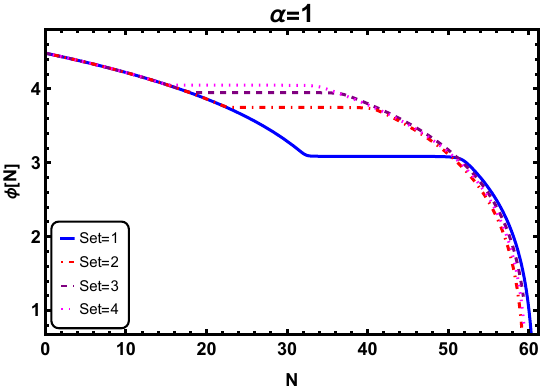}}
    \subfigure[\label{a1_bgn_H}]{\includegraphics[width=7.2cm]{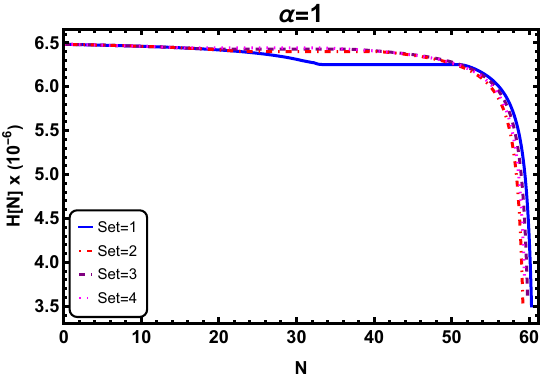}}
    \subfigure[\label{a1_bgn_e1}]{\includegraphics[width=7.2cm]{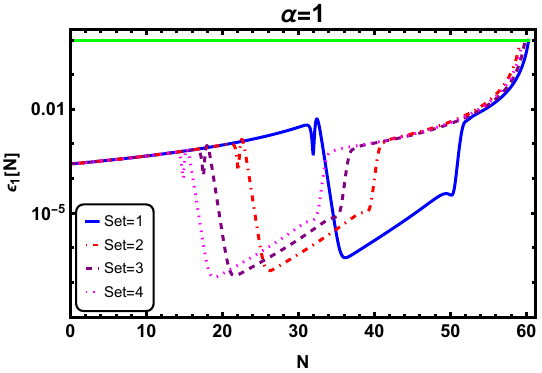}}
    \subfigure[\label{a1_bgn_e2}]{\includegraphics[width=7cm]{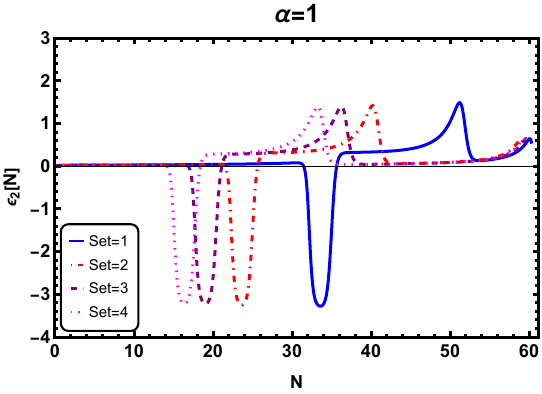}}
    \caption{The plot shows the scalar field, Hubble parameter, slow-roll parameter $\epsilon_1$, and $\epsilon_2$ versus the number of e-fold for $\alpha = 1$ and four sets of $(\phi_c, \xi_0, \xi_1)$ parameters. In addition, $V_0$ is taken as $V_0 = 1.29 \times 10^{-10}$, the scalar field at the crossing time is taken as $\phi_\star = 4.482$. }
    \label{a1_bgn}
\end{figure}

In the first case, the $\alpha$ parameter of the Mutated Hilltop potential is taken as $\alpha = 1$. The solutions for the Hubble parameter, scalar field, and the slow-roll parameters $\epsilon_i$ are presented versus the number of e-folds in Fig.\ref{a1_bgn} for four sets of the $(\phi_c, \xi_0, \xi_1)$ parameters. 
The evolution of the scalar field and the Hubble parameter mimic their behavior in the slow-roll regime, except when near the fixed point. Around this point, their evolution contains a flat regime where the velocity of the scalar field decreases significantly, and we have a USR regime. One could find the same behavior for the slow-roll parameters. The first slow-roll parameter $\epsilon_1$ experiences a dip in the vicinity of the fixed point where its magnitude suddenly drops. It touches the order of the magnitude $10^{-7}$, then it is expected to have an enhancement in the resulting power spectrum. Then, $\epsilon_1$ increases and reaches $\epsilon = 1$ around the number of e-fold $N = 60$, where inflation ends. The second slow-roll parameter $\epsilon_2$ has USR behavior near the fixed point, where the slow-roll approximation is no longer valid. The bench mark values of $\phi_c$, $\xi_0$, and $\xi_1$ are chosen such that the total duration of inflation lasts around $60$ e-folds by taking the same initial field $\phi_\star = 4.482$ at $N = 0$. It is realized that by taking the higher value of $\phi_c$, a bigger value of the parameter $\xi_1$ is required to end the inflationary phase. In addition, the dip in the slow-roll parameter $\epsilon_1$ occurs at smaller values of number of e-fold by increasing $\phi_c$.  \\ 
\begin{table}[h]
    \centering
    \begin{tabular}{p{1.2cm}p{1cm}p{1cm}p{2.5cm}p{1.3cm}p{2cm}}
    \hline \\[-0.2cm]
            & $\ \phi_c$ & $\xi_1$ & $\qquad \ \xi_0$ & $\mathcal{P_{\zeta}}^{\rm peak}$ & $\quad k^{\rm peak}$ \\[0.2cm]
    \hline \\[-0.2cm]
        set 1 & $3.095$ & $28.1$ & $-2.27228 \times 10^7$ & $0.0331$ & $2.084 \times 10^{13}$ \\[0.2cm]
        set 2 & $3.750$ & $54.5$ & $-5.53702 \times 10^6$ & $0.0414$ & $1.070 \times 10^{9}$ \\[0.2cm]
        set 3 & $3.950$ & $66.0$ & $-3.67733 \times 10^6$ & $0.0485$ & $1.081 \times 10^{7}$ \\[0.2cm]
        set 4 & $4.050$ & $73.0$ & $-2.98612 \times 10^6$ & $0.0482$ & $8.044 \times 10^{5}$ \\
    \hline
    \end{tabular}
    \caption{The table shows the values of the free parameters of the model for all four sets and the maximum value of the power spectrum.}
    \label{table_a1}
\end{table}

\begin{figure}
    \centering
    \includegraphics[width=8cm]{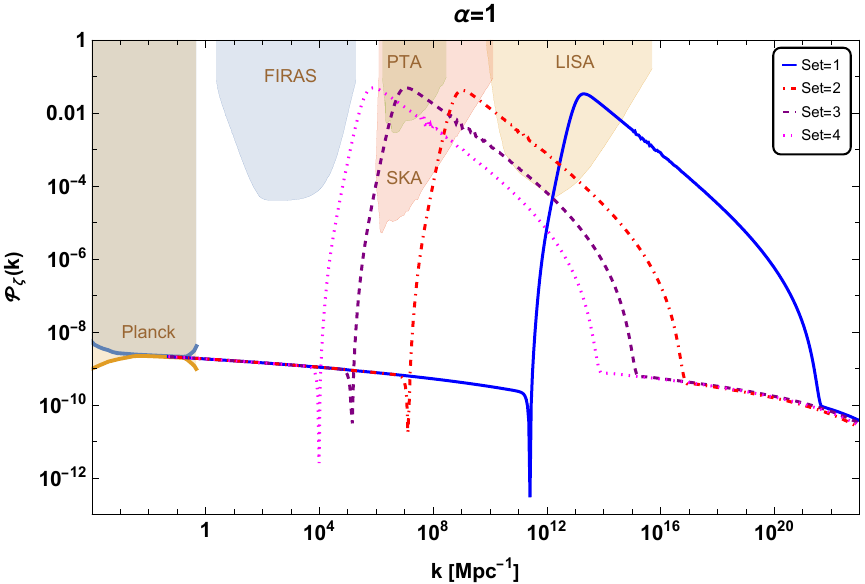}
    \caption{The power spectrum versus the mode $k$ is plotted for $\alpha = 1$ and different sets of $(\phi_c, \xi_0, \xi_1)$ parameters with $V_0 = 1.29 \times 10^{-10}$ and $\phi_\star = 4.482$. The resulting power spectrum at the pivot scale is obtained as $\mathcal{P}_{\zeta}(k_\star) = 2.098 \times 10^{-9}$. The scalar spectral index and the tensor to the scalar ratio are estimated as $n_s = 0.973$ and $r = 0.004$.}
    \label{ps_a1}
\end{figure}

\noindent The scalar power spectrum is obtained by substituting the numerical solution of the Mukhanov-Sasski equation in Eq.\eqref{Ps}. The resulting power spectrum is illustrated in Fig.\ref{ps_a1} for different sets of $(\phi_c, \xi_0, \xi_1)$ parameters. It starts with a scale-invariant regime, then increases, reaches the order of the $\mathcal{O}(0.01)$, and decreases. The peak of the enhancement occurs at a different scale for each set. The result is placed in Table.\ref{table_a1}, where one can find more numerical information about the values of the parameters and the magnitude of the power spectrum. It is realized that by increasing the free parameter $\phi_c$, the peak of the power spectrum is placed in the large scale $k$. This will result in a higher magnitude for the mass of PBH. \\ 
The numerical estimated value for the scalar power spectrum at the pivot scale $k_\star = 0.05 {\rm Mpc^{-1}}$ is about $2.098 \times 10^{-9}$, which is good agreement with the Planck-2018~\cite{Planck:2018jri}. The scalar spectral index is around $n_s = 0.973$, and the tensor-to-scalar ratio is $r = 0.004$.  \\ 

\subsection{$\alpha = 2$}
\begin{figure}[t]
    \centering
    \subfigure[\label{a1_bgn_phi}]{\includegraphics[width=7cm]{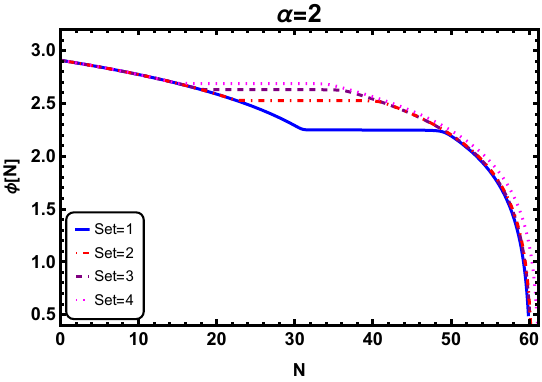}}
    \subfigure[\label{a1_bgn_H}]{\includegraphics[width=7cm]{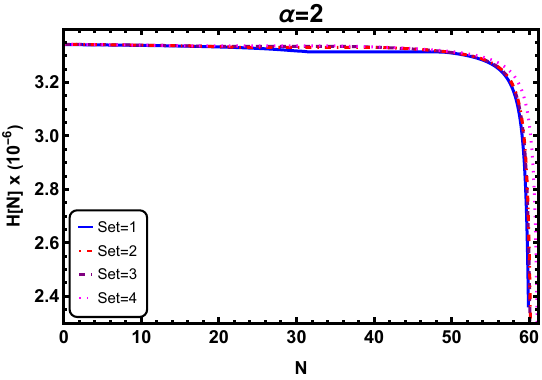}}
    \subfigure[\label{a1_bgn_e1}]{\includegraphics[width=7.2cm]{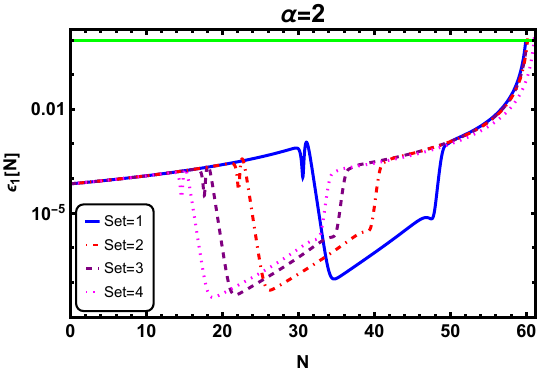}}
    \subfigure[\label{a1_bgn_e2}]{\includegraphics[width=7cm]{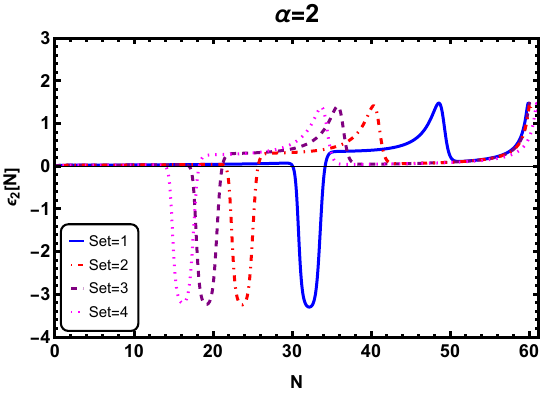}}
    \caption{The plot shows the scalar field, Hubble parameter, slow-roll parameter $\epsilon_1$, and $\epsilon_2$ versus the number of e-fold for $\alpha = 2$ and four sets of $(\phi_c, \xi_0, \xi_1)$ parameters. In addition, $V_0$ is taken as $V_0 = 3.372 \times 10^{-11}$, the scalar field at the crossing time is taken as $\phi_\star = 2.91$. }
    \label{a2_bgn}
\end{figure}

Setting $\alpha = 2$, the background solutions are obtained from Eqs.\eqref{Friedmann}-\eqref{eom}. Using the solution, the evolution of the scalar field, Hubble parameter, and the slow-roll parameters $\epsilon_i$ are obtained, depicted in Fig.\ref{a2_bgn}. 
The flat regime in the evolution of $\phi$ and $H$ implies the USR regime where the velocity of the field is reduced dramatically. The system is near the fixed point during this regime, and the scalar field remains almost unchanged. The slow-roll parameter $\epsilon_1$ undergoes a rapid reduction at first, reaching an order of $10^{-7}$, then increasing. The inflationary phase ends as $\epsilon = 1$ that occurs at the number of e-fold around $N = 60$. The behavior of the slow-roll parameter $\epsilon_2$ indicates that the slow-roll approximation is violated during the USR. Taking the initial value of the scalar field $\phi_\star = 2.91$ at $N = 0$, the parameters are set to produce about $N = 60$ number of e-fold for all cases, refer to Table.\ref{table_a1}. It is found that by increasing the parameter $\phi_c$, higher values of $\xi_1$ are required to fix the total number of e-fold of the inflationary phase. Also, the parameter $\xi_0$ is determined through Eq.\eqref{x0_constraint}. In addition, the dip in the slow-roll parameter $\epsilon_1$ occurs earlier by amplifying the $\phi_c$ parameter, meaning the USR regime starts earlier.     \\ 
\begin{table}[h]
    \centering
    \begin{tabular}{p{1.2cm}p{1cm}p{1cm}p{2.5cm}p{1.3cm}p{2cm}}
    \hline \\[-0.2cm]
            & $\phi_c$ & $\ \xi_1$ & $\qquad \ \xi_0$ & $\mathcal{P_{\zeta}}^{\rm peak}$ & $\quad k^{\rm peak}$ \\[0.2cm]
    \hline \\[-0.2cm]
        set 1 & $2.250$ & $61.3$ & $-1.69423 \times 10^7$ & $0.0348$ & $5.285 \times 10^{12}$ \\[0.2cm]
        set 2 & $2.529$ & $104.0$ & $-5.59988 \times 10^6$ & $0.0407$ & $1.194 \times 10^9$ \\[0.2cm]
        set 3 & $2.632$ & $127.0$ & $-3.71384 \times 10^6$ & $0.0485$ & $1.201 \times 10^7$ \\[0.2cm]
        set 4 & $2.690$ & $140.5$ & $-2.98039 \times 10^6$ & $0.0480$ & $6.617 \times 10^5$ \\
    \hline
    \end{tabular}
    \caption{The table shows the values of the free parameters of the model for all four sets and the maximum value of the power spectrum.}
    \label{table_a2}
\end{table}

\begin{figure}
    \centering
    \includegraphics[width=8cm]{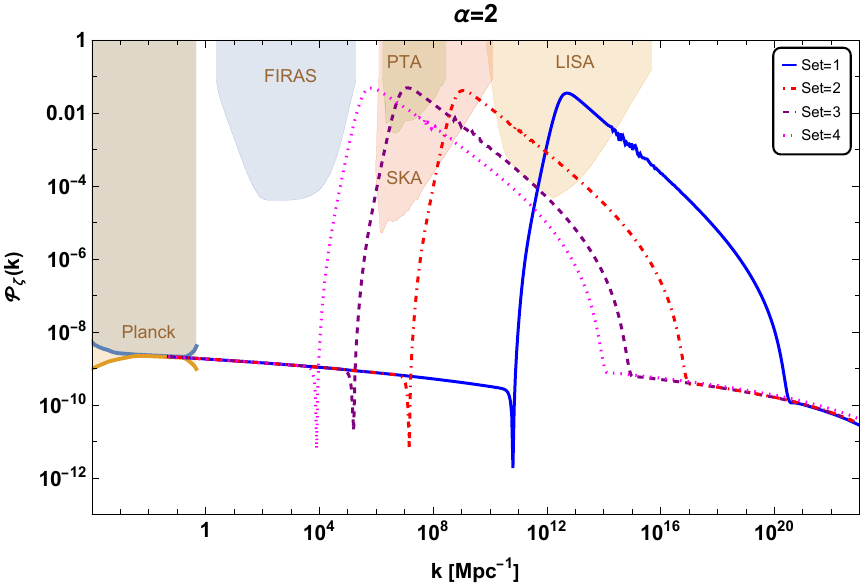}
    \caption{The power spectrum versus the mode $k$ is plotted for $\alpha = 2$ and different sets of $(\phi_c, \xi_0, \xi_1)$ parameters with $V_0 = 3.372 \times 10^{-11}$ and $\phi_\star = 2.91$. The resulting power spectrum at the pivot scale is obtained as $\mathcal{P}_{\zeta}(k_\star) = 2.098 \times 10^{-9}$. The scalar spectral index and the tensor to the scalar ratio are estimated as $n_s = 0.973$ and $r = 0.004$.}
    \label{ps_a2}
\end{figure}

\noindent Feeding the Mukhanov-Sasaki equation with the background solution, the scalar power spectrum is calculated using Eq.\eqref{Ps}. Fig.\ref{ps_a2} displays the resulting power spectrum versus the wavenumber $k$ for different sets of $(\phi_c, \xi_0, \xi_1)$ parameters. For all sets, the power spectrum has scale-invariant behavior for small $k$. Then, it increases and achieves the magnitude of the order of $\mathcal{O}(0.01)$. This enhancement in the power spectrum is directly related to the dip in the slow-roll parameter $\epsilon_1$. The peak in $\mathcal{P}_{\zeta}$ occurs at different values of $k$, which is reported in Table.\ref{table_a2}. The results indicate that by increasing the parameter $\phi_c$, the peak in the power spectrum happens at smaller mode $k$ or at larger scale. This is the result that we expect to have based on the behavior of $\epsilon_1$. As the dip occurs in a smaller number of e-fold, the USR happens earlier in the inflationary phase. Therefore, the enhancement of the power spectrum at this time is related to the perturbations with smaller mode $k$. 

\section{Analysis of PBH Formation}\label{numerical_pbh_formation}
In this section, we will apply the prescription mentioned in sec.~\ref{analysis} to calculate the mass of produced PBHs and corresponding mass function $\psi(M)$ and DM fraction $f_{PBH}$ in $1/3< \omega \leq 1$ regime. As we have already discussed, PBH formation requires the enhancement in the primordial power spectrum ($\mathcal{P_{\zeta}}(k)$). For this, we take Mutated Hilltop as our working model in the EGB background. During the calculation of the abundance, the most important quantity one needs to calculate is the mass fraction ($\beta(M)$). Integrating the Eq.\ref{eq:beta}, we find
\begin{equation}
    \beta(M) =\rm erfc \left( \frac{\delta_c}{\sqrt{2} \sigma(M)} \right)
    \label{beta1}
\end{equation}
Where "erfc" is the complementary error function and $\delta_c$ is the critical density, above which any over-dense region would collapse to form a PBH. Since 1974, several attempts have been made to calculate the $\delta_c$ precisely. Carr and Hawking made the original estimation of the $\delta_c$, utilizing the Newtonian gravity's Jeans instability criteria to infer $\delta_c \sim c^2_s$, in case of a static fluid $ c^2_s = \omega$, for radiation $\omega =1/3 = \delta_c$. Since then, several methods have been adopted in the literature to estimate the $\delta_c$, including the hydrodynamical solutions, numerical relativity and more~\cite{Niemeyer:1997mt,1978SvA....22..129N,1979ApJ...232..670B,1980SvA....24..147N,shibata99,Hawke:2002rf}. However, for our analysis we will follow the analytic expression obtained by \textit{Harada et.al} in~\cite{Harada:2013epa} where authors have adopted a three-zone model of the overdensity profile to obtain $\omega$ depend expression of the $\delta_c$ \footnote{The shape of the density profile can alter $\delta_c$ significantly, for detailed discussion readers are suggested to go through~\cite{Musco:2018rwt,Kalaja:2019uju,Germani:2018jgr}} as:
\begin{equation}
\delta_c = \frac{3 \left(1+\omega \right)}{\left( 5+ 3~\omega \right)} \sin^2 \left( \frac{\pi \sqrt{\omega}}{\left( 1+3~\omega \right)} \right)
\label{deltac}
\end{equation}
It is noteworthy to mention that Eq.~\ref{deltac} is not valid for the PBHs forming in Matter Dominated (MD) epoch (see~\cite{Harada:2016mhb} for PBH formation in MD.).

\subsection{$\alpha=1$ Case}
\begin{figure}[t]
    \centering
    \subfigure[\label{mpsim1}]{\includegraphics[width=7cm]{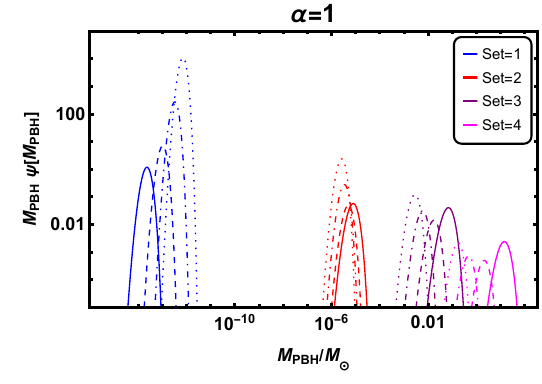}}
    \caption{The plot shows the evolution of $M_{PBH} \psi(M_{PBH})$ with the PBH mass $M_{PBH}$ for different sets. The color Blue, Purple, Red and Magenta signifies the different sets mentioned in Table~\ref{table_a1}. The solid line is for $\omega=1/3$, dashed line is for $\omega=0.5$, dot-dashed line is for $\omega=0.7$ and dotted line represents $\omega=0.9$ }
    \label{mpsimplot_a1}
\end{figure}

Equipped with the prerequisites, we can proceed to the calculation of mass function and DM fraction for our model. To calculate the variance of the density perturbations, we numerically solve the Eq.~\ref{sigmaM1} with $P_{\zeta}(k)$ shown in Fig.~\ref{ps_a1} using the Gaussian window function and adopting a $\omega$ dependent transfer function. Once we obtain the solution of Eq.~\ref{sigmaM1}, it is straight forward to calculate the $\psi(M)$ using Eqs.~\ref{MpsiMw}, ~\ref{beta1} with Eq.~\ref{deltac}. 
\begin{figure}[t]
    \centering
    \subfigure[\label{mpsimTa1s1}]{\includegraphics[width=7cm]{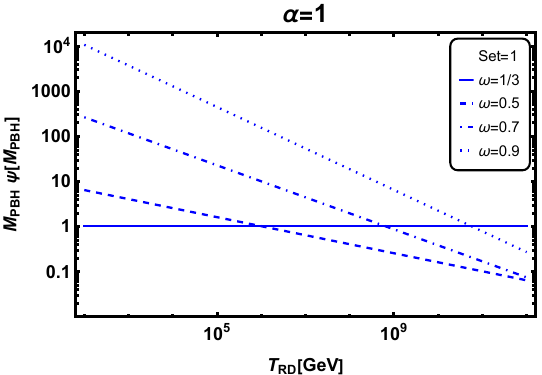}}
    \subfigure[\label{mpsimTa1s2}]{\includegraphics[width=7cm]{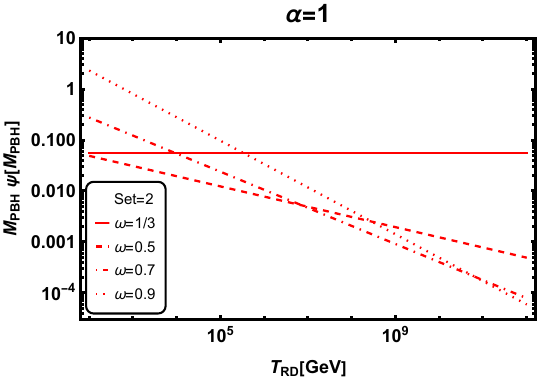}}
    \subfigure[\label{mpsimTa1s3}]{\includegraphics[width=7.2cm]{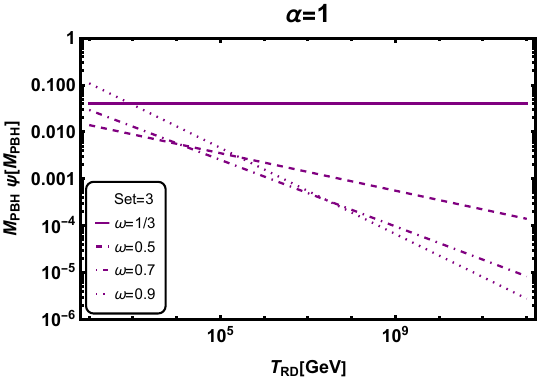}}
    \subfigure[\label{mpsimTa1s4}]{\includegraphics[width=7cm]{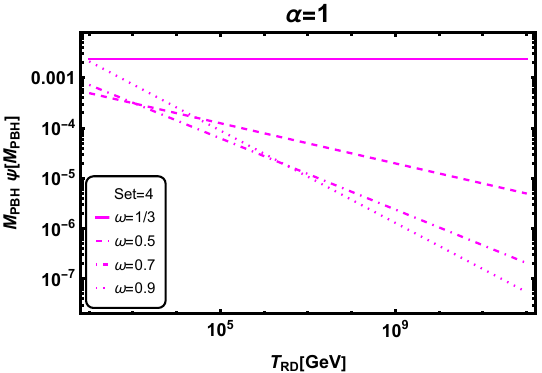}}
    \caption{The plot shows behavior of $M_{PBH} \psi(M_{PBH})$ with $T_{RD}$ for different sets. Plot legends are self-explanatory. }
    \label{mpsimT_a1}
\end{figure}
It is evident from the Eq.~\ref{MpsiMw} that $\psi(M)$ crucially depends on the $\omega$ and $T_{RD}$. In our analysis, we take four different values of $\omega=(1/3, 0.5, 0.7, 0.9)$ and $T_{RD}=100 \rm Gev$. However, it is only meaningful to calculate the quantity $M \psi(M)$. The evolution of $M \psi(M)$ versus $M_{PBH}$ for different EoS $\omega$ is demonstrated in Fig.~\ref{mpsimplot_a1}. Here we have consider only single value of $T_{RD}= (100 \rm Gev )$, however $M_{PBH} \psi(M_{PBH})$ is dependent on $T_{RD}$. We have shown this dependency in Fig.~ \ref{mpsimT_a1}.

\subsection{$\alpha=2$ Case}
\begin{figure}[t]
    \centering
    \subfigure[\label{mpsimTa2s1}]{\includegraphics[width=7cm]{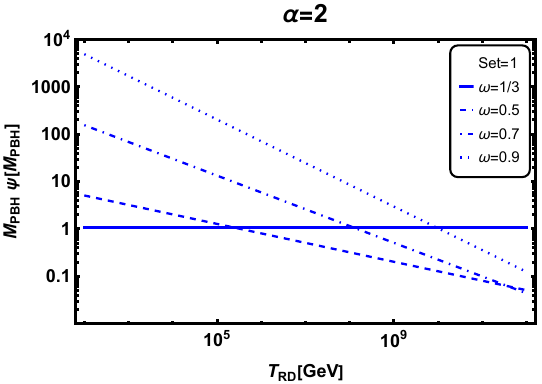}}
    \subfigure[\label{mpsimTa2s2}]{\includegraphics[width=7cm]{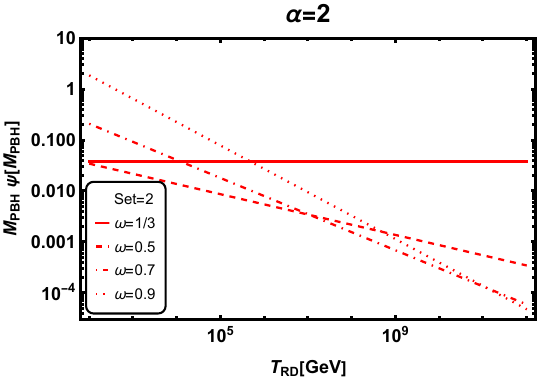}}
    \subfigure[\label{mpsimTa2s3}]{\includegraphics[width=7.2cm]{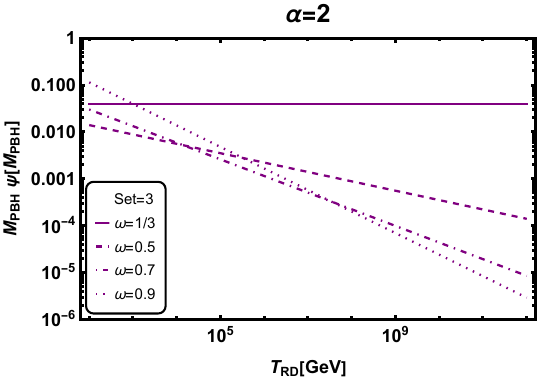}}
    \subfigure[\label{mpsimTa2s4}]{\includegraphics[width=7cm]{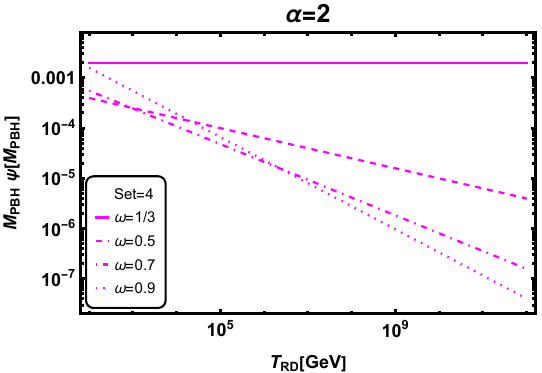}}
    \caption{The plot shows behavior of $M_{PBH} \psi(M_{PBH})$ with $T_{RD}$ for different sets. Plot legends are self-explanatory. }
    \label{mpsimT_a2}
\end{figure}

\begin{figure}[t]
    \centering
    \subfigure[\label{mpsim2}]{\includegraphics[width=7cm]{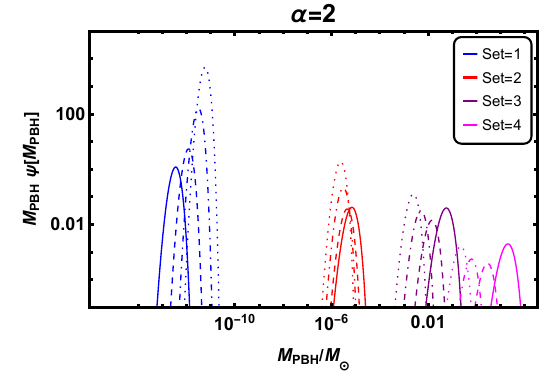}}
    \caption{The plot shows the evolution of $M_{PBH} \psi(M_{PBH})$ with the PBH mass $M_{PBH}$ for different sets. The color Blue, Purple, Red and Magenta signifies the different sets mentioned in Table~\ref{table_a2}. The solid line is for $\omega=1/3$, dashed line is for $\omega=0.5$, dot-dashed line is for $\omega=0.7$ and dotted line represents $\omega=0.9$ }
    \label{mpsimplot_a2}
\end{figure}

Similar to the $\alpha=1$ case, we carry out the required steps to obtain the $M_{PBH} \psi(M_{PBH})$ with produced PBHs mass $M_{PBH}$ for different sets and $\omega$; see Fig.~\ref{mpsimplot_a2}. Behavior of $M_{PBH} \psi(M_{PBH})$ versus $T_{RD}$ can found in Fig.~\ref{mpsimT_a2}.  

\begin{figure}[t]
    \centering
    \subfigure[\label{fpbha1}]{\includegraphics[width=7cm]{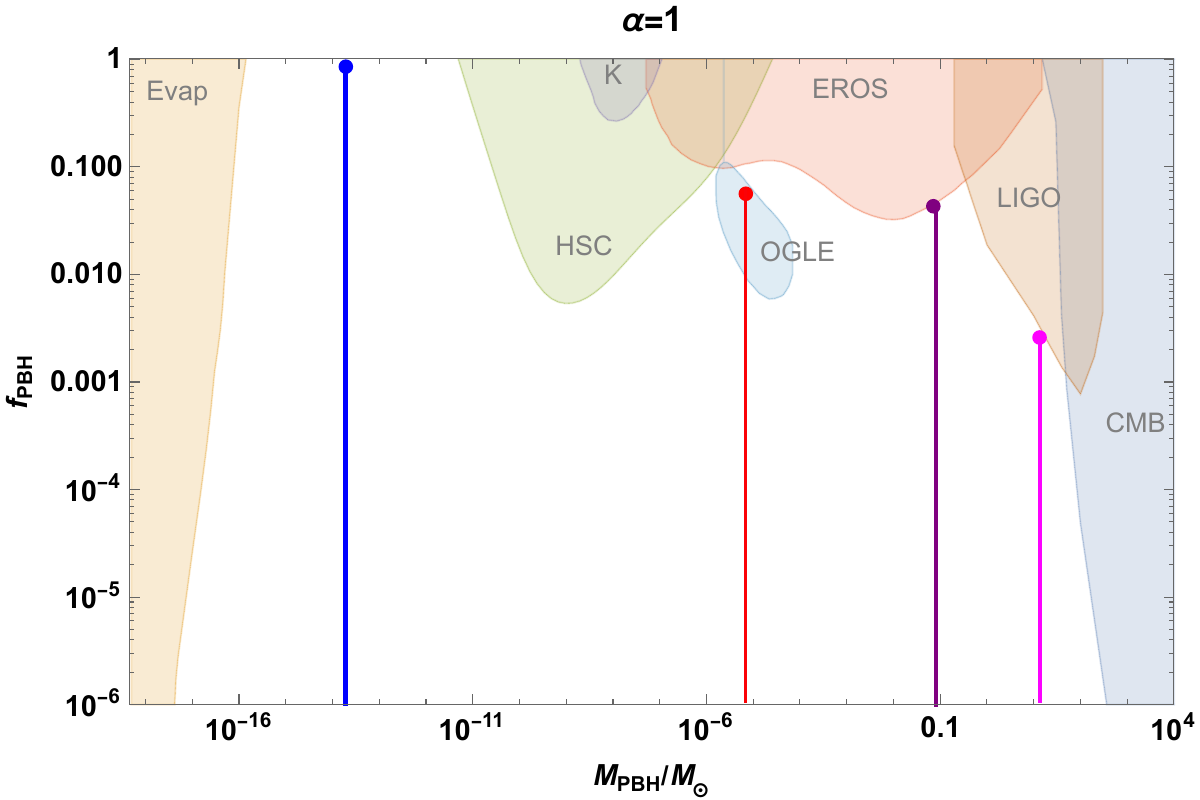}}
    \subfigure[\label{fpbha2}]{\includegraphics[width=7cm]{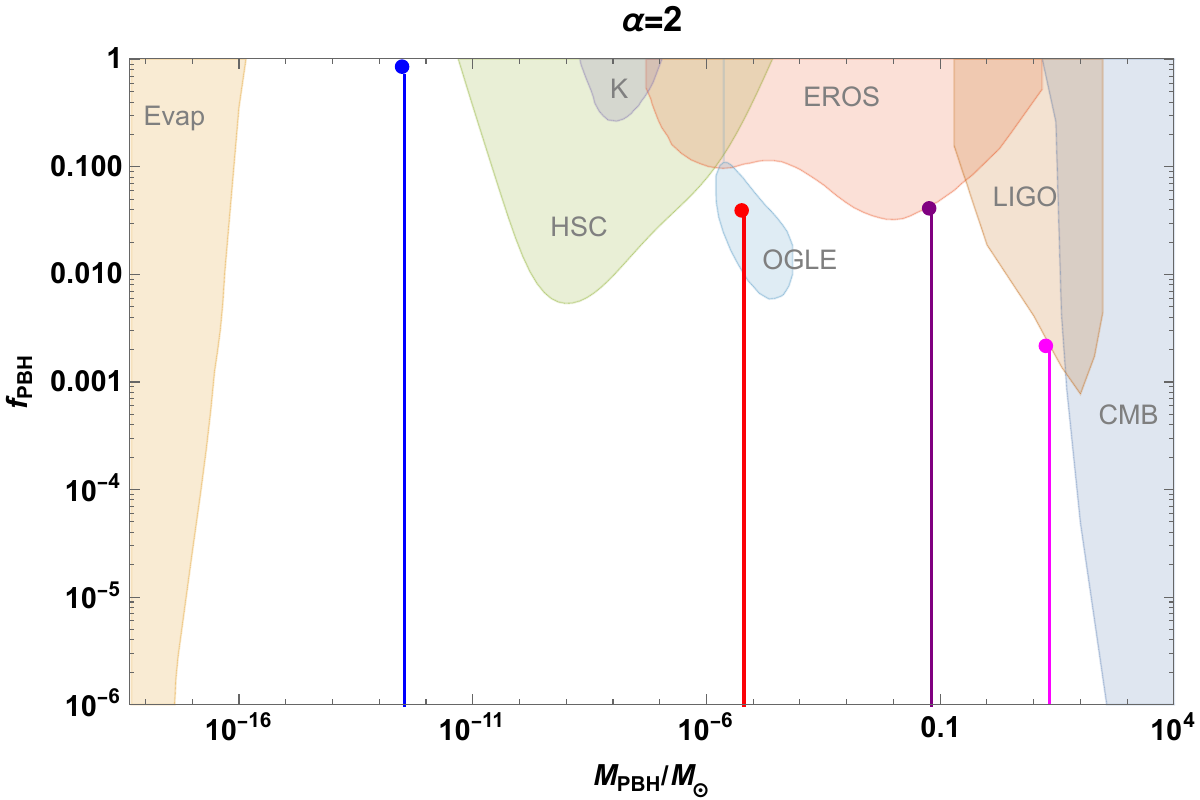}}
    \caption{The plot shows the behavior of $f_{PBH}$ with $M_{PBH}$ for the case where PBH formed in radiation-dominated epoch ($\omega=1/3$) for two different values of $\alpha= (1,2)$ and different colors represents different sets. The color blue is for set-1, color red is for set-2, purple color denotes set-3, and color magenta represents set-4. The observation bounds on PBH abundance are plotted using~\cite{bradley_j_kavanagh_2019_3538999}
    \label{fpbhplot}. ``Evap" represents the bound on PBH evaporation~\cite{Lehmann:2018ejc} through Hawking evaporation, HSC~\cite{Niikura:2017zjd}, Kepler(k)~\cite{Griest:2013aaa}, EROS~\cite{Tisserand:2006zx}, OGLE~\cite{Niikura:2019kqi}, LIGO~\cite{Inomata:2018epa} and  ‘CMB’ denotes bounds from Compton drag and Compton cooling of CMB photons~\cite{Ali-Haimoud:2016mbv} }
\end{figure}

The total DM fraction ($f_{PBH}$) for all the mentioned sets can be evaluated from Eq.~\ref{ftot}. The PBHs analysis for the both $\alpha=(1,2)$ appears identical (see Figs~\ref{mpsimplot_a1},\ref{mpsimplot_a2}). This is because we choose the benchmark point for $(\phi_c, \xi_0, \xi_1)$ such that the allowed $ f_{PBH}$ does not violet the current observation bounds at least for the PBHs which are forming during the radiation dominated epoch ($\omega=1/3$). However, the total DM fraction ($f_{PBH}$) can be larger or smaller than 1 for the PBHs forming in an epoch where $\omega \neq 1/3$ for the same set. The reason behind this non-trivial behavior of $f_{PBH}$, or rather we say $M_{PBH} \psi(M_{PBH})$ is twofold. First, for the over-densities to collapse in different epochs, it is required to cross different values of critical threshold $\delta_c$, which can be clearly seen from Eq.~\ref{deltac}. This will lead to a different value of $\beta(M)$ even for the same $P_{\zeta}(k)$, which results in different $\psi(M_{PBH})$. Secondly, the mass of the produced PBHs ($M_{PBH}$) also depends on the $k$ and $\omega$ (see Eq.~\ref{Mkexact}). So, for the same $P_{\zeta}(k)$, the produced mass can be different for different values of $\omega$. This non-trivial behavior of $M_{PBH} \psi(M_{PBH})$ is being demonstrated in Figs~\ref{mpsimplot_a1} and \ref{mpsimplot_a2}, where for the same set (take set-1 for instance) the amplitude of  $M_{PBH} \psi(M_{PBH})$ is different for distinct values of $\omega$. This will lead to the different $f_{PBH}$ value. Although from Eq.~\ref{ftot} it is evident that $f_{PBH}$ would be number but for the sake of comparison with the various observational bounds we show $f_{PBH}$ in Fig.~\ref{fpbhplot} for the PBHs which formed during radiation epoch ($\omega=1/3$). The exact value of the $f_{PBH}$ would differ depending on the $\omega$ even for the same set of parameters. Exact values of the $M_{PBH}$ and their corresponding $f_{PBH}$ for the PBHs formed in different epochs can be found in Table~\ref{table_fpbh_a1} and ~\ref{table_fpbh_a2}. It is demonstrated in Table~\ref{table_fpbh_a1} and ~\ref{table_fpbh_a2} that for the same set of parameters, the abundance of PBHs can be much larger or smaller depending on the $\omega$. \\ 

The data from numerous types of observation have put severe constraints on the $f_{PBH}$, PBHs with $M_{PBH}< 2.5 \times 10^{-19}  M_\odot$ have been evaporated completely through the Hawking radiation, and they not be a DM candidate\cite{Hawking:1974rv}. However, slightly heavier PBHs can evaporate $\gamma-$rays, photons, neutrinos, gravitons and other particles. Observations from extra-galactic radiation background, Voyager, SPI/INTEGRAL can constrained the PBHs with $M_{PBH} \leq   10^{-17} M_\odot$~\cite{Siegert:2016ijv,Laha:2019ssq,Super-Kamiokande:2011lwo,Dasgupta:2019cae,Laha:2020ivk}. Energy injection by the particles generated through PBH evaporation can alter the CMB anisotropies and abundance of light elements in BBN this can constrain $M_{PBH} \geq 5.5 \times 10^{-21} M_\odot $ and  $  M_{PBH} \approx 10^{-22}-10^{-21} M_\odot $; whereas PBHs in mass range $10^{-16 } M_\odot \leq M_{PBH} \leq 5 \times 10^{-12} M_\odot $ posses the most intriguing possibility of explaining the $100 \%$ DM density~\cite{Carr:2016drx,Carr:2020xqk,Inomata:2017okj,Ballesteros:2017fsr,Bertone:2016nfn}.
PBHs, falling in the mass range $10^{-11 } M_\odot \leq M_{PBH} \leq 10^{-1} M_\odot $, are constrained by their gravitational lensing. Observations from HSC~\cite{Niikura:2017zjd}, EROS~\cite{Tisserand:2006zx} and OGLE~\cite{Niikura:2019kqi} have ruled out the PBHs contribution to DM above $1-10 \%$~\cite{Smyth:2019whb,Tisserand:2006zx,Niikura:2017zjd,Niikura:2019kqi}. Also, the abundance of PBHs in the mass region $0.2 M_\odot \leq M_{PBH} \leq 300 M_\odot $ are tightly constrained by LIGO/Virgo collaborations~\cite{Ali-Haimoud:2017rtz,Bird:2016dcv,Sasaki:2016jop,Cholis:2016kqi,Clesse:2016vqa,DeLuca:2020qqa}. Lastly, CMB spectrum and anisotropies can be impacted by the radiation of accreted gas by PBHs of mass $  M_{PBH} \geq 100 M_\odot $~\cite{carr1981pregalactic,Ricotti:2007au,Serpico:2020ehh}. In Fig.~\ref{fpbhplot}, we show some of these observational bounds along with the $f_{PBH}$ obtained in our model.

\begin{table}
\begin{tabular}{|p{1cm}|p{2.2cm}p{1.2cm}|p{2.2cm}p{1.2cm}|p{2.2cm}p{1.2cm}|p{2.2cm}p{1.2cm}|}
      \hline
      & \multicolumn{2}{|c|}{set-1} & \multicolumn{2}{|c|}{set-2}  & \multicolumn{2}{|c|}{set-3}  & \multicolumn{2}{|c|}{set-4} \\
     \hline 
     \rule{0pt}{3ex} 
     $\omega$ & $\ M_{\rm PBH}/M_\odot$ & $f_{\rm PBH}$ & $\ M_{\rm PBH}/M_\odot$ & $f_{\rm PBH}$ & $\ M_{\rm PBH}/M_\odot$ & $f_{\rm PBH}$ & $\ M_{\rm PBH}/M_\odot$ & $f_{\rm PBH}$ \\
     \hline
     \rule{0pt}{3ex} 
     $1/3$ & $1.91 \times 10^{-14}$ & $0.9973$ & $0.0711$ & $0.0441$ & $7.24 \times 10^{-6}$ & $0.0551$ & $12.8408$ & $0.0026$ \\
     $0.5$ & $9.08 \times 10^{-14}$ & $5.2984$ & $0.0186$ & $0.0140$ & $4.77 \times 10^{-6}$ & $0.0424$ & $2.0074$ & $0.0004$ \\
     $0.7$ & $3.03 \times 10^{-13}$ & $196.13$ & $0.0066$ & $0.0289$ & $3.45 \times 10^{-6}$ & $0.2275$ & $0.4771$ & $0.0006$ \\
     $0.9$ & $6.86 \times 10^{-13}$ & $7575.07$ & $0.0033$ & $0.1106$ & $2.78 \times 10^{-6}$ & $1.9001$ & $0.1807$ & $0.0019$ \\
     \hline 
\end{tabular}
\caption{Table for $\alpha=1$, for various sets, this table shows the explicit values PBHs mass $M_{PBH}$ produced in different epoch $\omega$ and their corresponding values of $f_{PBH}$. }
\label{table_fpbh_a1}
\end{table}

\begin{table}
\begin{tabular}{|p{1cm}|p{2.2cm}p{1.2cm}|p{2.2cm}p{1.2cm}|p{2.2cm}p{1.2cm}|p{2.2cm}p{1.2cm}|}
      \hline
      & \multicolumn{2}{|c|}{set-1} & \multicolumn{2}{|c|}{set-2}  & \multicolumn{2}{|c|}{set-3}  & \multicolumn{2}{|c|}{set-4} \\
     \hline 
     \rule{0pt}{3ex} 
     $\omega$ & $\ M_{\rm PBH}/M_\odot$ & $f_{\rm PBH}$ & $\ M_{\rm PBH}/M_\odot$ & $f_{\rm PBH}$ & $\ M_{\rm PBH}/M_\odot$ & $f_{\rm PBH}$ & $\ M_{\rm PBH}/M_\odot$ & $f_{\rm PBH}$ \\
     \hline
     \rule{0pt}{3ex} 
     $1/3$ & $2.97 \times 10^{-13}$ & $1.00$ & $5.82 \times 10^{-6}$ & $0.0402$ & $0.0575$ & $0.0420$ &  $18.977$ & $0.0020$ \\
     
     $0.5$ & $1.07 \times 10^{-12}$ & $4.0930$ & $3.92 \times 10^{-6}$ & $0.0313$ & $0.0154$ & $0.0136$ & $2.8531$ & $0.0003$ \\
     
     $0.7$ & $2.90 \times 10^{-12}$ & $112.69$ &  $2.88 \times 10^{-6}$ & $0.0055$ & $0.0287$ & $0.1733$ & $0.6580$ & $0.0005$ \\
     
     $0.9$ & $5.67 \times 10^{-12}$ & $3402.79$ & $2.35 \times 10^{-6}$ & $1.4969$ & $0.0028$ & $0.1117$ & $0.2442$ & $0.0015$ \\
     \hline 
\end{tabular}
\caption{Table for $\alpha=2$, for various sets, this table shows the explicit values PBHs mass $M_{PBH}$ produced in different epoch $\omega$ and their corresponding values of $f_{PBH}$. }
\label{table_fpbh_a2}
\end{table}

\section{Gravitational waves}\label{numerical_GW}
The produced cosmological perturbations are the source of the background gravitational waves. At the linear order of the perturbations, the scalar and tensor perturbations are decoupled and evolve independently. Assuming no anisotropic stresses, the tensor perturbations evolve source-free at the first order of perturbations, and the corresponding power spectrum is given by \cite{Caprini:2018mtu,Christensen:2018iqi}
\begin{equation}
    \Delta_h(k, \eta) = T(k, \eta) \; \mathcal{P}_{h, {\rm inf}}(k)
\end{equation}
in which $\mathcal{P}_{h, {\rm inf}}(k)$ is the primordial tensor power spectrum at the horizon re-entry, 
\begin{equation}
    \mathcal{P}_{h, {\rm inf}}(k) = \frac{2}{\pi^2} \; \frac{H^2_\star}{M_p^2} \; \left( \frac{k}{k_\star} \right)^{n_t}
\end{equation}
where $k_\star$ is the pivot scale, $H_\star$ stands for the Hubble parameter when the pivot scale exits the horizon during the inflationary phase, and $n_t$ is the spectral index. $T(k, \eta)$ is the transfer function that depends on the background evolution from when a mode $k$ re-enters the horizon until it is observed. Assuming instant reheating, in standard cosmology, a radiation-dominant phase starts after the inflationary phase, followed by a matter-dominant phase. However, before the radiation-dominant phase, the universe may go through a non-standard epoch with EoS that differs from ones in the radiation- and matter-dominated phases. Therefore, a distinct transfer function for the re-entering modes is expected. In such non-standard thermal history of the universe and up to the first order of the perturbations, the energy density of the produced GW is acquired as \cite{Figueroa:2019paj,Bernal:2019lpc,Bernal:2020ywq}
\begin{equation}
\Omega_{\rm GW,0}^{(1)}(k) = \frac{\Omega_{\rm rad, 0}}{12\pi^2}\bigg(\frac{g_{*,k}}{g_{s,k}}\bigg)\bigg(\frac{g_{s,0}}{g_{s,k}}\bigg)^{4/3}\bigg(\frac{H_\star}{M_{\rm Pl}}\bigg)^2 \; \frac{\Gamma ^2 (\alpha +1/2)}{2^{2(1-\alpha}\alpha ^{2\alpha}\Gamma ^2(3/2)}\mathcal{W}(\kappa)\kappa^{2(1-\alpha)},\label{GW_1_gen}
\end{equation}
where $\alpha=\frac{2}{1+3w}$, $\kappa=\frac{k}{k(T_1)}=\frac{f}{f(T_1)}$, $T_1$ is the temperature at the start of the radiation-dominated phase, and
\begin{equation}
\mathcal{W}(\kappa)= \frac{\pi \alpha}{2\kappa}\bigg[\bigg(\kappa J_{\alpha+1/2}(\kappa)-J_{\alpha-1/2}(\kappa)\bigg)^2+\kappa^2J^2 _{\alpha-1/2}(\kappa)\bigg],\label{Bessel_GW1}
\end{equation}
where $J_{i}$ is the Bessel function of order $i$. \\ 
The scalar and tensor perturbations are coupled at the second order of perturbations, and their evolution is not independent. The second-order tensor perturbations are fed by the first-order perturbations so that they are evolved as \cite{Espinosa:2018eve,Kohri:2018awv}
\begin{equation}
    h_k'' + 2 {\cal H} h_k' + k^2 h_k=  {\cal S}(\mathbf{k}, \eta)\, ,
\end{equation}
where ${\cal S}(\mathbf{k}, \eta)$ has the role of the source for the second-order tensor perturbation that depends on the first-order perturbations. Due to the enhancement of the scalar power spectrum, induced GWs are expected to be produced that might be large enough to meet the sensitivity of the current and future GW detectors. \\ 
Here, we also consider the possibility that the universe undergoes a non-standard era and might not be in the radiation-dominant phase after inflation. Therefore, the induced GW can be generated during a non-standard era. Assuming that the observed GW signal is the second-order GW induced by the first-order scalar perturbations, we investigate the effects of standard and non-standard post-inflationary on the generated induced GWs~\cite{Domenech:2024rks,Liu:2023hpw,Liu:2023pau,Chen:2024roo}. The resulting energy density parameter from the induced GW has been investigated in \cite{Ananda:2006af,Baumann:2007zm,Kohri:2018awv,Domenech:2021ztg,Domenech:2019quo,Domenech:2020kqm,Witkowski:2022mtg,Balaji:2023ehk,Domenech:2021wkk}, obtained to be
\begin{equation}\label{induced_GW}
    \Omega_{GW}^w(k) = \mathcal{N} \, {\left(\frac{k}{k_{\rm rh}}\right)^{-2b}} \int_0^1 \textrm{d} d \int_1^\infty \textrm{d} s \, I_\omega(d,s) \, \mathcal{P}_\zeta \bigg(\frac{k}{2}(s+d)\bigg) \mathcal{P}_\zeta \bigg(\frac{k}{2}(s-d)\bigg) \, ,
\end{equation}
where $b \equiv (1-3w)(1+3w)$, and the integration kernel $I_\omega(d, s)$ is given by 
\begin{align}\label{I_omega}    
I_\omega(d,s) = & \hphantom{\times} \mathcal{F}(b) {\bigg(\frac{(d^2-1)(s^2-1)}{d^2-s^2}\bigg)}^2 \, 
\frac{|1-y^2|^b }{(s^2-d^2)^2} \, \\ \nonumber 
 &\hspace{3cm} \times \Bigg\{\hphantom{+} \hphantom{\frac{4}{\pi^2}} \bigg[\mathsf{P}_{b}^{-b}(y) + \frac{2+b}{1+b} \mathsf{P}_{b+2}^{-b}(y) \bigg]^2 \Theta \big(s-c_s^{-1} \big) \\\nonumber 
 &\hspace{3.5cm} \hphantom{\times \Bigg\{} +  \frac{4}{\pi^2} \bigg[\mathsf{Q}_{b}^{-b}(y) + \frac{2+b}{1+b} \mathsf{Q}_{b+2}^{-b}(y) \bigg]^2 \Theta \big(s-c_s^{-1} \big) \\ \nonumber 
 & \hspace{3.5cm} \hphantom{\times \Bigg\{} +  \frac{4}{\pi^2} \bigg[\mathcal{Q}_{b}^{-b}(-y) + 2 \frac{2+b}{1+b} \mathcal{Q}_{b+2}^{-b}(-y) \bigg]^2 \Theta \big(c_s^{-1}-s \big) \Bigg\} \, ,
\end{align}
where the propagation speed of the scalar perturbations is indicated by $c_s$. $\mathsf{P}^{\mu}_{\nu}(x)$ and $\mathsf{Q}^{\mu}_{\nu}(x)$ stand for the Ferrers function of the first and second kinds. $\mathcal{Q}^{\mu}_{\nu}(x)$ is the associated Legendre function of the second kind, and $y$ and $\mathcal{F}_b$ are defined as
\begin{equation}
    y \equiv \frac{s^2+d^2-{2}{c_s^{-2}}}{s^2-d^2} \, , \quad 
    \mathcal{F}(b)=\frac{1}{3}\left( \frac{4^{1+b}(b+2)}{\left(1+b\right)^{1+b}(2b+3)c_s^2} \, \Gamma^2 \Big[b+\tfrac{3}{2} \Big] \right)^2 .
\end{equation}
\begin{figure}[t]
    \centering
    \subfigure[\label{sigw_a1}]{\includegraphics[width=13cm]{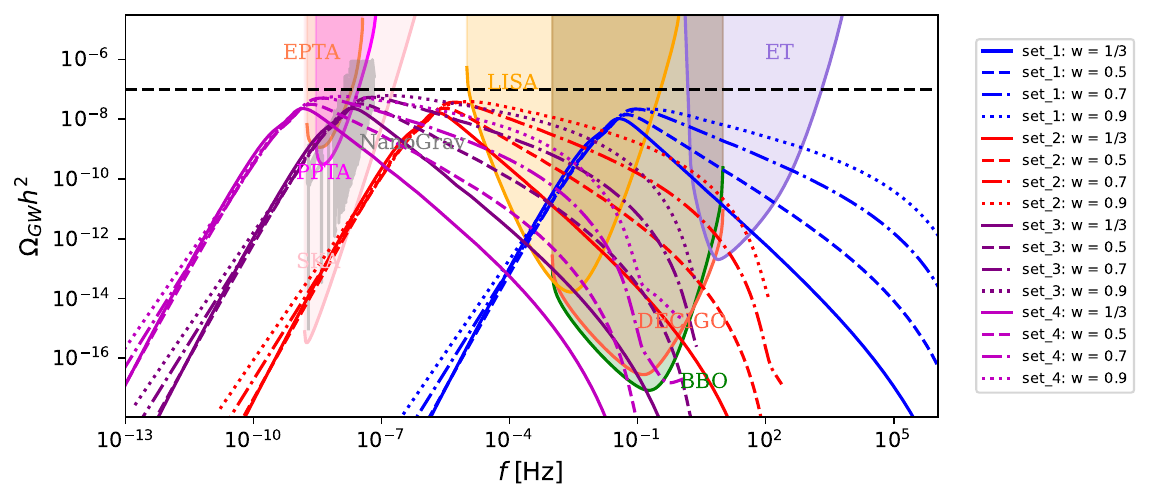}}
    \subfigure[\label{sigw_a2}]{\includegraphics[width=13cm]{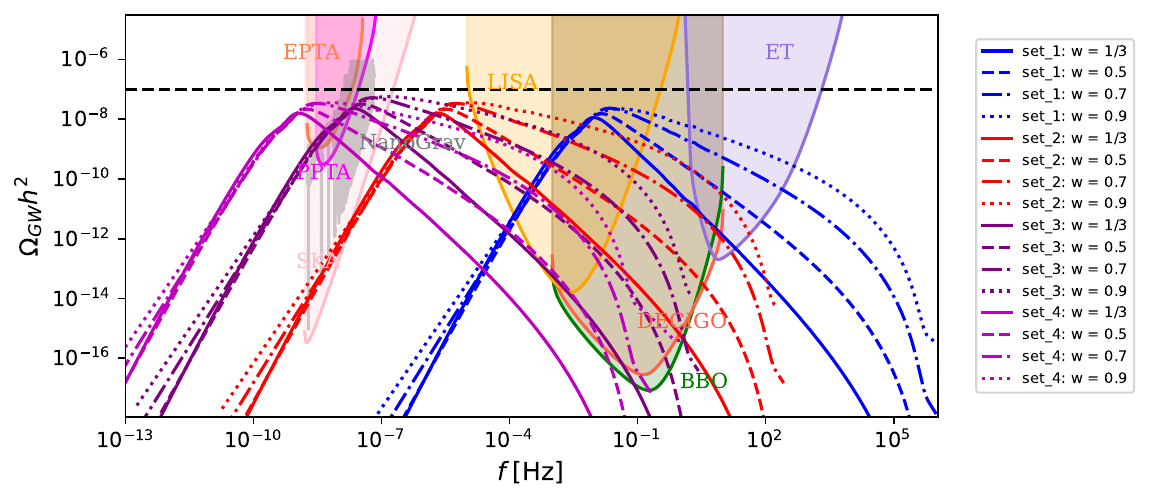}}
    \caption{The plot shows the produced induced GW for a)$\alpha = 1$, and b)$\alpha = 2$ for different sets of parameter and distinct values of EoS parameter $\omega$. }
    \label{sigw}
\end{figure}
The obtained induced GWs for $\alpha = 1$ and $\alpha = 2$ are displayed in Figs.\ref{sigw_a1} and \ref{sigw_a2}, respectively, for different sets of parameters and different values of $\omega$\footnote{For GW calculation we have used the publicly available code \texttt{SIGWfast}~\cite{Witkowski:2022mtg}}. The peak of GWs is located at different frequencies for different sets of parameters, and its magnitude is of the order of $10^{-8}$ for all cases. The peak of the induced GWs for the first set of parameters stands in $10^{-2} \; {\rm Hz}$. The calculated induced GW signal for this set finely crosses the sensitivity bound of future experiments, such as 
LISA~\cite{Danzmann:1997hm,Harry:2010zz,LIGOScientific:2014qfs,LISA:2017pwj,LISACosmologyWorkingGroup:2022jok}, 
BBO~\cite{Crowder:2005nr,Corbin:2005ny,Harry:2006fi,Yagi:2011wg,Yagi:2017wg}, 
DECIGO~\cite{Seto:2001qf,Kawamura:2011zz,Kawamura:2006up}, and 
ET~\cite{Punturo_2010,Branchesi:2023mws}. For the second set, the obtained induced GW signal crosses the LISA, BBO, and DECIGO sensitivity bound. Also, the ET experiment can measure the signal by assuming a non-standard era with $\omega > 0.7$. The peak of the induced GWs for the third set of parameters is placed in $10^{-8} \; {\rm Hz}$ frequency, which stays in the observatory range of SKA \cite{Carilli:2004nx,Moore:2014lga,Weltman:2018zrl,Correa:2023whf}, and also it can explain the NANOGrav result~\cite{NANOGrav:2023gor,NANOGrav:2023hde,Gangopadhyay:2023qjr}. The induced GW signal for the last set of parameters crosses the SKA observatory bound. From Fig\ref{sigw}, it is realized that in the non-standard era, with EoS $1/3 \leq \omega <1$, the peak in the induced GWs is enhanced by increasing $\omega$.

\section{Conclusion}\label{conclusion}
We investigated the formation and abundance of the PBHs in EGB gravity theory, where the scalar field has a coupling with the EGB term. The formation of PBH required a significant enhancement in the scalar power spectrum during the inflationary phase. Such enhancement is usually supported by adding a manual term to the potential in the canonical scalar field model or by imposing a suitable coupling term in modified gravity theories. It is realized that the inflationary phase is first in the slow-roll regime and then transits to the USR regime. In this regime, the velocity of the scalar field reduces dramatically and the field remains almost constant. Due to this, there is a dip in the slow-roll parameter $\epsilon_1 = -\dot{H}/H^2$, where it decreases significantly. However, the second slow-roll parameter can exceed the magnitude one, and the slow-roll approximations are violated. The universe remains in the USR regime for some number of e-folds, then it transits back to the SR regime and reaches the end of the inflationary phase. 

We considered the potential to be a Mutated Hilltop one with the GB coupling term as in \eqref{coupling}. After choosing the suitable values of coupling the parameters, the inflationary phase was successfully generated with about $60$ e-fold. The results of the power spectrum at the pivot scale, scalar spectral index, and tensor-to-scalar ratio were examined to agree with the CMB measurement~\cite{Planck:2018jri}. To consider the model's flexibility, four sets of parameters were established so that for each set, the enhancement in the power spectrum starts at a different number of e-folds, and its peak occurs at different scales. 

After the end of inflation, as the perturbations re-enter the horizon, the large over-densities gravitationally collapse, leading to the formation of PBHs. Depending on scales of power spectrum enhancement, PBH with wide mass range, $< \mathcal{O} 10 {\rm M_\odot}$ are formed in the post-inflationary era. The enhanced perturbations may re-enter the horizon in a non-standard era, such as a stiff-fluid dominant era, that can severely affect the mass and abundance of PBHs and the induced GWs. Therefore, we generalized our study and investigated the effects of a non-standard thermal history on the mass of PBHs, their abundance, and the produced induced GWs, for all sets of parameters. As mentioned, we take  Mutated Hilltop potential as our working model. Two distinct values of the potential parameter $\alpha =(1,2)$ were considered, and corresponding to each value of $\alpha$, we had four sets.\\
For the $\alpha=1$ case, we choose four different values of coupling parameters ($\xi_0, \xi_1$ and $\phi_c$); see Table~\ref{table_a1}. Different values of these parameters will lead to different evolutions of the background quantities and power spectrum. As power spectrum is the main quantity that enters into the calculation of PBHs and GW, different sets result in different $M_{PBH}$ Eq.~\ref{Mkexact} and $\psi_{M_{PBH}}$ Eq.~\ref{MpsiMw} both of these quantities are the function of $\omega$ and $T_{RD}$. In Fig.~\ref{mpsimplot_a1} we show the explicit dependency of $\omega$ on $M_{PBH} \psi (M_{PBH})$ for potential parameter $\alpha=1$ and constant value of $T_{RD} = (100 \rm Gev)$ for different sets consider here. If one notices there is linear behavior of $M_{PBH} \psi (M_{PBH})$ with $M_{PBH}$, for set-1 $M_{PBH} \psi (M_{PBH})$ increases with the increasing values of $\omega$, however in other sets (2,3, and 4) there is no linear behavior. The reason behind such no-trivial behavior of $M_{PBH} \psi (M_{PBH})$ is because for some values of $k_{peak}$ ($k$ value where $\mathcal{P}_s$ is maximum ) the $M_{PBH}$ increasing with the $\omega$ where some other $k_{peak}$ the mass decreases with $\omega$ (see Eq. \ref{Mkexact}). Secondly, the critical density $\delta_c$ is also a function of EoS, so $\delta_c$ can be different for different EoS. Due to these two reasons, the quantity $M_{PBH} \psi (M_{PBH})$ does not show a linear behavior with $\omega$. This will result in the different PBH abundance $f_{PBH}$ (see Table ~\ref {table_fpbh_a1}). For $\omega>1/3$ in set-1, the $f_{PBH}$ goes beyond 1, which seems unphysical. It means the PBHs forming in a non-standard epoch would require a smaller amplitude of $\mathcal{P}_\zeta^{\rm peak}$ than the PBHs forming in radiation. Whereas for other sets, the situation might differ depending on the $\omega$. Furthermore different $T_{RD}$ could also leads to different abundances, in Fig.\ref{mpsimT_a1} we demonstrate the effect of $T_{RD}$ on $M_{PBH} \psi (M_{PBH})$. For $\alpha=2$ we found a similar behavior of $M_{PBH} \psi (M_{PBH})$ and $f_{PBH}$ with the changing $\omega$ and $T_{RD}$ (see Fig. \ref{mpsimplot_a2} and \ref{mpsimT_a2}).  In both the cases $\alpha=(1,2)$, we choose the benchmark points such that in radiation, we have a set (set-1) where PBHs can be attributed to $100 \%$ DM density ($f_{PBH}=1$)

The tensor perturbations are coupled to the scalar perturbations in the second order. Then, the scalar-induced GW is expected to be generated. The induced GWs for two values of $\alpha$ and all sets of parameters were calculated, and the results are exhibited in Fig.\ref{sigw}, where we generalized our consideration by including the non-standard epoch. The peak of the produced induced GWs, for the first sets of parameters, is placed at the $10^{-1} \; {\rm Hz}$ frequency, which stands in the future experiment observatory bounds of LISA, BBO, and DECIGO. The signal of this induced GW signal crosses the ET observatory bound as well. For the second set of parameters, the peak of the induced GWs is at the frequency of about $10^{-6} \; {\rm Hz}$, and its signal crosses the bounds of LISA, BBO, and DECIGO observatories. The NANOGrav results can be explained by the model with the third set of parameters. For this set, the peak of the GW is at the frequency $10^{-8} \; {\rm Hz}$, and the signal also crosses the SKA boundary range. The induced GWs signal for the last set of the parameter crosses the SKA bound so that the peak stands on the range $10^{-9} \; {\rm Hz}$. It also can explain the NANOGrav results. Our results indicate that by including the non-standard epoch, the peak of the produced induced GWs rises, and the corresponding signal can explain the current and future observations. In our future work, we would like to explore the combined effect of Non-Gaussanity and EoS on the production of PBHs and GW~\cite{Pi:2024jwt,Firouzjahi:2023xke,Franciolini:2023pbf,Ferrante:2022mui,Choudhury:2023kdb,Choudhury:2023fwk,Ragavendra:2023ret}.

\section*{Acknowledgments}
A.M. would like to thank A. Ashrafzadeh and K. Karami for the fruitful discussion on the EGB gravity. Yogesh would like to thank S.  Bhattacharya for helping develop the \texttt{Mathematica} code for the PBH formation in the non-standard era. The authors would also like to thank M. R. Gangopadhyay for the useful discussions. 


\bibliographystyle{apsrev4-1}
\bibliography{MH_EGB.bib}




\end{document}